\documentclass{article}
\usepackage{graphicx} 
\usepackage[numbers]{natbib}  
\usepackage{hyperref}         
\usepackage{amsmath} 
\usepackage{authblk} 
\usepackage{xcolor}
\usepackage{url}
\usepackage{booktabs}
\usepackage[a4paper, margin=1in]{geometry}

\title{Gender Bias in Perception of Human Managers Extends to AI Managers} 

\author[1, 2, 3, *]{Hao Cui}
\author[1, 4, 5]{Taha Yasseri}
\affil[1]{School of Social Sciences and Philosophy, Trinity College Dublin, Dublin, Ireland}
\affil[2]{School of Sociology, University College Dublin, Dublin, Ireland}
\affil[3]{Geary Institute for Public Policy, University College Dublin, Dublin, Ireland}
\affil[4]{Faculty of Arts and Humanities, Technological University Dublin, Dublin, Ireland}
\affil[5]{School of Mathematics and Statistics, University College Dublin, Dublin, Ireland} 
\affil[*]{Corresponding author: cuih1@tcd.ie}

\begin{document}

\maketitle

\begin{abstract}
As AI becomes more embedded in workplaces, it is shifting from a tool for efficiency to an active force in organizational decision-making. Whether due to anthropomorphism or intentional design choices, people often assign human-like qualities, including gender, to AI systems. However, how AI managers are perceived in comparison to human managers and how gender influences these perceptions remains uncertain. To investigate this, we conducted randomized controlled trials (RCTs) where teams of three participants worked together under a randomly assigned manager. The manager was either a human or an AI and was presented as male, female, or gender-unspecified. The manager's role was to select the best-performing team member for an additional award. Our findings reveal that while participants initially showed no strong preference based on manager type or gender, their perceptions changed \textcolor{black}{notably} after experiencing the award process. As expected, those who received awards rated their managers as more trustworthy, competent, and fair, and they were more willing to work with similar managers in the future. In contrast, those who were not selected viewed them less favorably. However, male managers, whether human or AI, were more positively received by awarded participants, whereas female managers, especially female AI managers, faced greater skepticism and negative judgments when they did not give awards. These results suggest that gender bias in leadership extends beyond human managers to include AI-driven decision-makers as well. As AI assumes more managerial responsibilities, understanding and addressing these biases will be crucial for designing fair and effective AI management systems.
\end{abstract}


\section{Introduction}

The increasing integration of artificial intelligence (AI) in workplaces is transforming its function from merely enhancing efficiency to actively shaping organizational decision-making, sparking discussions on the evolving dynamics between humans and machines~\cite{tsvetkova2024new}. AI algorithms are increasingly taking on roles traditionally performed by humans, such as assisting with decisions related to promotion and hiring~\cite{kelan2024algorithmic, mainka2019algorithm}. While human managers often retain final authority, some companies are experimenting with AI in leadership-adjacent roles, such as managing repetitive tasks or optimizing team operations~\cite{kiron2023workforce}. The growing influence of AI raises questions about workplace dynamics, particularly how AI decision-makers are perceived relative to human managers. Issues such as fairness, biases, organizational justice~\cite{colquitt2001justice}, trust, morale, and willingness to collaborate within teams are becoming increasingly important as AI reshapes the future of leadership and management. 


\textcolor{black}{This study aims to examine how the combined effects of manager type (human or AI), manager gender, self-perceived contribution, and award outcome influence the perceptions of managers' trustworthiness, competence, fairness, and participants' willingness to work with a similar manager in future tasks. Trustworthiness relates to reliability and ethical standing; competence underpins a manager's authority and efficacy; fairness is crucial for morale and acceptance; and willingness to work in future endeavors reflects the sustainability of the collaboration. These dimensions are essential for fostering positive and effective human-AI team interactions. By examining these dynamics, this study offers insights into how gendered presentations in AI managers impact human perceptions, ultimately contributing to a deeper understanding of effective AI integration in the workplace. The following section reviews prior research that informs our hypotheses.}

\section{\textcolor{black}{Background}}

\subsection{\textcolor{black}{Gender bias in leadership evaluations}}

Gender bias often influences how people assess a manager's fairness, competence, and leadership qualities, and it has been widely studied. Studies consistently reveal that female managers are often perceived less favorably than their male counterparts~\cite{heilman1995sex, heilman2001description, koch2015meta, braun2017think}, potentially due to the discrepancy between the traditional female gender role and the leadership role~\cite{elsesser2011does, eagly2002role}. Women in managerial positions are often held to higher standards of warmth and likability~\cite{ridgeway2001gender}, while also being scrutinized more harshly for assertiveness or decisiveness -- traits typically valued in male managers~\cite{rudman2008backlash}. Male managers are frequently perceived as more competent and authoritative by default~\cite{schein1996think, powell2002gender}, which can enhance perceptions of their fairness and leadership capability. Conversely, female managers, despite displaying the same level of competence, may face skepticism regarding their authority or be seen as less credible~\cite{eagly2002role}. Past research indicates that this disparity is especially pronounced in workplace evaluations by male peers~\cite{szymanska2018gender}, where female managers are rated significantly lower than their male counterparts, while female peers provide similar evaluations for both male and female managers~\cite{szymanska2018gender}. 

\subsection{\textcolor{black}{Anthropomorphism and gendering in AI systems}}

Studies indicate that people may have different expectations and biases toward AI versus human managers~\cite{logg2019algorithm, logg2024simple}. Research reveals that AI-driven decisions are often seen as objective and consistent~\cite{jones2023people}. AI's data-centric methods can reduce certain biases inherent in human decision-making~\cite{hofeditz2022applying}. However, AI may lack the empathy~\cite{montemayor2022principle} and contextual awareness~\cite{mitchell2021abstraction} expected from human managers, which can impact trust and acceptance. Research shows people may perceive AI algorithms as more unfair because they fail to incorporate certain qualitative information and context, which can undermine the belief of procedural fairness~\cite{newman2020eliminating}.  Other research shows participants may be more lenient in their evaluations of AI systems than human experts when they perceive the AI as having less control over unfavorable outcomes~\cite{jones2023people}.

Gender stereotypes and biases that affect human managers are likely to also influence perceptions of AI managers, as AI systems are often anthropomorphized and assigned gender markers such as voice, name, or avatar by designers and users~\cite{craiut2022technology, wong2023chatgpt}. Female AI systems are often perceived as warmer and human-like, making them more acceptable in healthcare contexts~\cite{borau2021most}. Research in human-robot interaction shows that female bots are frequently associated with communal qualities such as warmth, friendliness, and emotional sensitivity~\cite{gustavsson2005virtual, eyssel2012s, otterbacher2017s, stroessner2019social, borau2021most} -- traits traditionally attributed to women~\cite{eagly1984gender, ebert2014warm}. Studies further show that female service robots elicit greater satisfaction and positive experiences compared with male robots~\cite{seo2022female}, reflecting the persistence of traditional gender stereotypes in shaping perceptions of AI systems. However, gendered biases in AI do not always confer advantages. The assigned gender of an AI system affects human-AI cooperation, with research showing that female-labeled AI agents are more likely to be exploited than their human counterparts~\cite{bazazi2025ai}. This gendered bias extends to error mitigation scenarios, where male users preferred apologetic feminine voice assistants over masculine ones~\cite{mahmood2024gender}. These patterns suggest that bots and voice assistants not only reflect but may also reinforce and amplify existing gender biases~\cite{kiron2023workforce}.  

Even when AI lacks explicit gender markers, individuals tend to subconsciously assign gender based on factors like language style, voice, and the perceived roles of AI. Research shows that minimal gender cues, such as vocal cues embedded in a machine, can evoke gender-based stereotypical responses~\cite{nass1997machines}. A study on ChatGPT shows it is more likely to be perceived as male than female due to its competence-oriented attributes, such as providing information or summarizing text~\cite{wong2023chatgpt}. AI systems designed for caregiving or support, such as digital assistants like Siri and Alexa, tend to be feminized through their anthropomorphization~\cite{costa2018conversing}, reflecting societal norms that associate caregiving and service roles with women~\cite{costa2019ai}. The subconscious gendering of AI can influence perceptions of fairness and team dynamics, making it important to understand these biases to promote positive interactions in human-AI collaborations.

\subsection{\textcolor{black}{Fairness, self-perception and award outcomes}}

In addition to external factors such as manager type and perceived gender, internal factors like self-perceived contribution can also shape perceptions of fairness in decision-making~\cite{guo2014neural}. According to equity theory, individuals evaluate fairness by assessing the balance between their contributions and rewards in comparison with those of their peers~\cite{adams1963towards}. Research demonstrates that individuals who view themselves as high contributors but are not chosen for recognition often report lower fairness ratings, driven by unmet expectations and perceived inequity~\cite{adams1965inequity}. This dynamic highlights the role of internal self-assessments in fairness evaluations.

\subsection{\textcolor{black}{Hypotheses}}

\textcolor{black}{Together, the reviewed literature highlights how biases linked to manager type and gender can shape perceptions of managerial decision-making. Yet, the interplay between these factors, particularly in competitive contexts where only a subset of individuals receive recognition, remains underexplored.}
While gender bias is a central concern of this research, it is part of a broader investigation into how multiple interacting factors influence managerial perceptions. 
\textcolor{black}{Building on prior literature and theoretical foundations, we propose the following hypotheses:}

\begin{itemize}
	\item \textbf{H1:} Participants who receive an award will rate their manager as more trustworthy, competent, and fair, and express greater willingness to work with a similar manager in the future than participants who do not receive an award.
	
	\item \textbf{H2:} Managers presented as male (both human and AI) will receive more favorable ratings than those presented as female, especially among participants who receive an award.
	
	\item \textbf{H3:} 
	Managers presented as female (both human and AI) will receive more negative ratings than those presented as male, especially among participants who do not receive an award.
	
	\item \textbf{H4:} Participants who perceive themselves as high contributors but do not receive an award will rate their manager as less fair, particularly when the manager is human.
	
	\item \textbf{H5:} There will be significant interaction effects between manager type, manager gender, award outcome, and self-perceived contribution on participants' ratings of trustworthiness, competence, fairness, and willingness to work with a similar manager in the future.
\end{itemize}

\section{Materials and Methods}

\textcolor{black}{To explore these dynamics, we conducted online experiments where participants worked in teams under managerial scenarios varying by manager type and gender.}
The survey questions used in the pre-treatment and post-treatment surveys, along with screenshots from the experiment, are provided in the Supporting Information. This research complies with University College Dublin (UCD) Human Research Ethics Regulations and Human Research Ethics Committee (HREC) guidelines for research involving human participants. The research study protocol has been approved by UCD Human Research Ethics (HS-LR-24-10-Cui-Yasseri). Informed consent was obtained from all human participants prior to the commencement of the experiment.

\subsection{Participants}
Participants were recruited from the Prolific platform (\url{https://www.prolific.com/}), selecting U.S.-based adults aged 18 and older for an online experiment exploring perceptions of decision-making in scenarios involving human and AI managers. The sample included N = 556 participants, with an average age of 37.4 years ($\pm$12.2). The sample consisted of approximately 48.7\% female, 48.4\% male, and 2.9\% identifying as non-binary or other genders. Participants reported various educational backgrounds, ranging from secondary education to postgraduate degrees. (See details in Supporting Information Fig. S15.) 

\subsection{Design and Measures}

\begin{figure}[htbp] 
	\centering
	\includegraphics[width=0.9\textwidth]{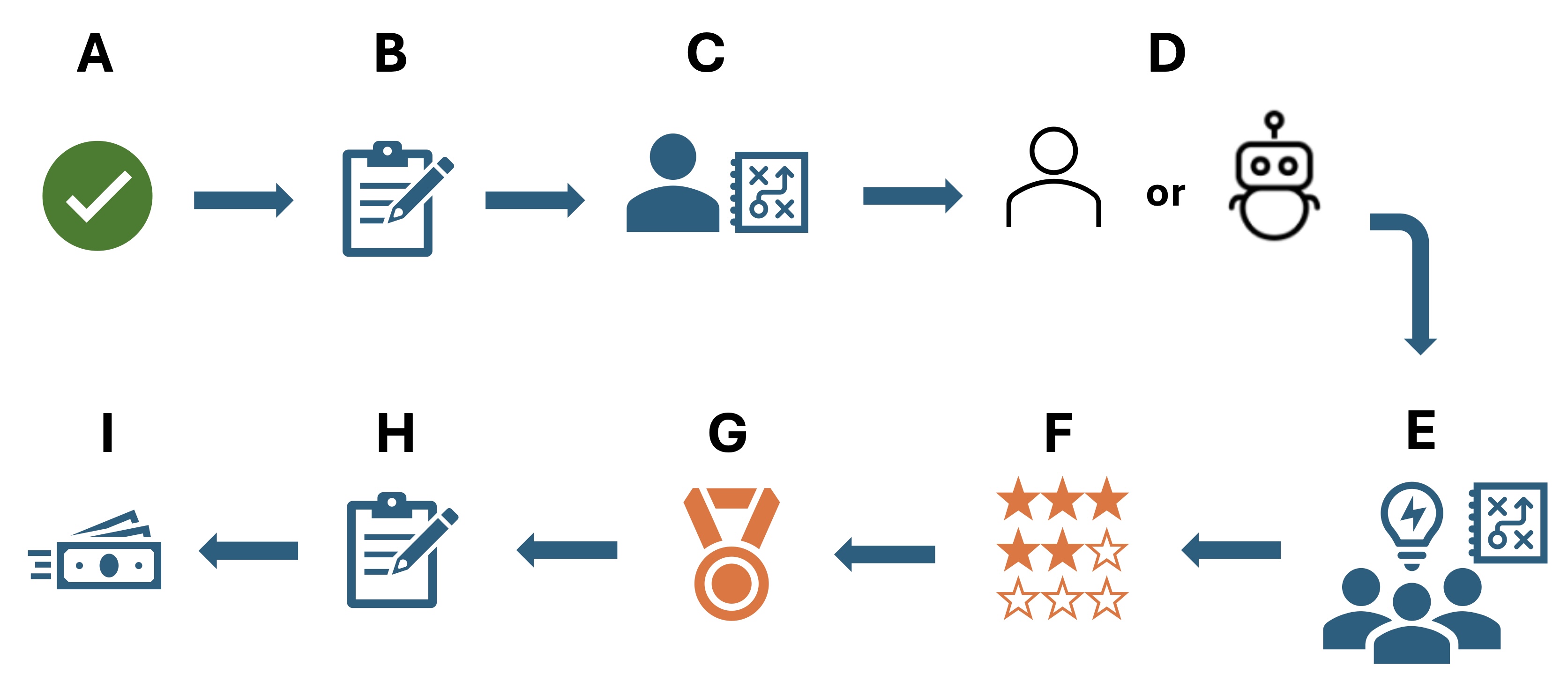} 
	\caption{Timeline of the experimental procedure. \textbf{(A)} Participant onboarding: participants read the information sheet and provide informed consent. \textbf{(B)} Pre-treatment survey. \textbf{(C)} Individual task. \textbf{(D)} Manager assignment: participants are informed they will work in teams of three in the next round. A manager is randomly assigned and introduced with type (AI or human) and gender (male, female, or unspecified). A representative image is shown for human or AI managers. Participants are told the manager will select the best-performing teammate based on problem-solving ability and communication skills, and the selected individual will receive a £0.50 extra award. 
		\textbf{(E)} Team task. \textbf{(F)} Contribution self-ratings. \textbf{(G)} Winner announcement. \textbf{(H)} Post-treatment survey. \textbf{(I)} Debrief and payment.}
	\label{fig:fig_game} 
\end{figure}

Figure~\ref{fig:fig_game} illustrates the timeline view of the experimental methodology.
This study employed a mixed experimental design to examine how participants' perceptions of trustworthiness, competence, fairness, and willingness to work with a manager were influenced by primary factors: manager type, manager gender, award outcome, and self-perceived contribution. Participants were randomly assigned to three-person groups, with each group placed in a different manager condition. In these conditions, participants viewed scenarios in which either a human or an AI manager made a selection decision. Within each manager type, the gender of the manager was also manipulated, with the manager identified as either female, male, or gender unspecified, to explore potential biases or differential expectations based on perceived gender. 

A key aspect of the design was the participants' award outcome. In each group, the assigned manager randomly chose a ``best player" to receive an extra award, and participants received an announcement about which participant had been selected as the ``best player". This manipulation allowed the study to investigate how participants' selection status interacted with both the manager type and manager gender, potentially shaping perceptions and attitudes toward the manager.

In addition to the award outcome, participants rated their own contributions to the teamwork as well as the contributions of their teammates on an 11-point scale ranging from 0 to 10. 
We define the contribution measure as

\begin{align*}
\text{Contribution} &= \text{Rating}_{\text{self}} - \frac{1}{2} \big( \text{Rating}_{\text{teammate 1}}  
+ \text{Rating}_{\text{teammate 2}} \big),
\end{align*} 
where $\text{Rating}_{\text{self}}$ is the rating an individual gives to themselves 
and $\text{Rating}_{\text{teammate i}}$  is the rating an individual gives to teammate i.
This self-assessed contribution measure captures how individuals perceive their own input relative to their peers. The rationale is that individual perceptions often depend not only on absolute judgments but on how people view themselves in relation to others. By comparing each player's self-rating with the average rating they assign to their teammates, we quantify the extent to which individuals perceive themselves as contributing more or less than others. This approach draws on theoretical frameworks from social comparison theory~\cite{festinger1957social}, which suggests that individuals evaluate themselves by comparing themselves to others, and with research on equity theory~\cite{adams1963towards, adams1965inequity}, where the perceived balance between one's contributions (inputs) and rewards (outputs) is crucial in shaping social judgments.

The dependent variables in this study were the participants' perceptions of trustworthiness, competence, fairness of their assigned manager, and the participants' willingness to collaborate with a similar manager in future tasks. Each of these perceptions was measured using a Likert-type scale from 0 to 10, with items such as ``How fair do you think the manager was in making the award decision in the game?".
Take fairness as an example; we employed the following linear regression model to examine the factors influencing post-treatment fairness perceptions:  
\begin{equation}
\begin{split}
\text{PF} = &\ \beta_0 + \beta_1 \text{PTF} + \beta_2 \text{PGF} + \beta_3 \text{A} + \beta_4 \text{E} + \beta_5 \text{G} + \notag \\
&\beta_6 (\text{AWO} \cdot \text{MT} \cdot \text{MG}) +
\beta_7 (\text{AWO} \cdot \text{MT} \cdot \text{C}) + \epsilon,
\label{eq:post_fairness}\\
\text{where}\\
\text{PF} & : \text{post-fairness (response)} \\
\text{PTF} & : \text{pre-type-fairness} \\
\text{PGF} & : \text{pre-gender-fairness} \\
\text{A}, \text{E}, \text{G} & : \text{Age, Education, and Gender} \\
\text{AWO} &: \text{Award Outcome, } \\
\text{MT}, \text{MG} &: \text{Manager Type}, \text{Manager Gender}\\
\text{C} &: \text{Contribution}\\ 
\beta_0 & : \text{Intercept term} \\
\beta_1, \beta_2, \ldots, \beta_7 & : \text{Coefficients for predictors and interactions} \\
\epsilon & : \text{Error term}
\end{split}
\end{equation}

The model includes pre-treatment fairness perceptions based on manager type (pre-type-fairness) and manager gender (pre-gender-fairness), along with demographic variables such as age, education, and gender. Interaction terms were included to capture the combined effects of awards, manager type, manager gender, and contribution on post-treatment fairness. Specifically, the model accounts for the three-way interaction between award outcome, manager type, and manager gender, as well as the three-way interaction between awards, manager type, and contribution.
The model was applied separately to female and male participants.

Overall, this design and these measures were intended to provide a detailed understanding of how variations in manager characteristics and selection outcomes influence key perceptions in decision-making contexts.

\subsection{\textcolor{black}{Experimental task}} 

\textbf{Experimental platform and infrastructure.} The experiments were conducted using Empirica (\url{https://empirica.ly/}), a platform designed for building and deploying interactive, real-time experiments~\cite{almaatouq2021empirica}. Empirica enables the creation of dynamic multiplayer tasks and facilitates participant interaction in controlled online environments. The experiments were run on the DigitalOcean droplet (\url{https://www.digitalocean.com/}).

\begin{figure}[htbp] 
	\centering
	\includegraphics[width=0.9\textwidth]{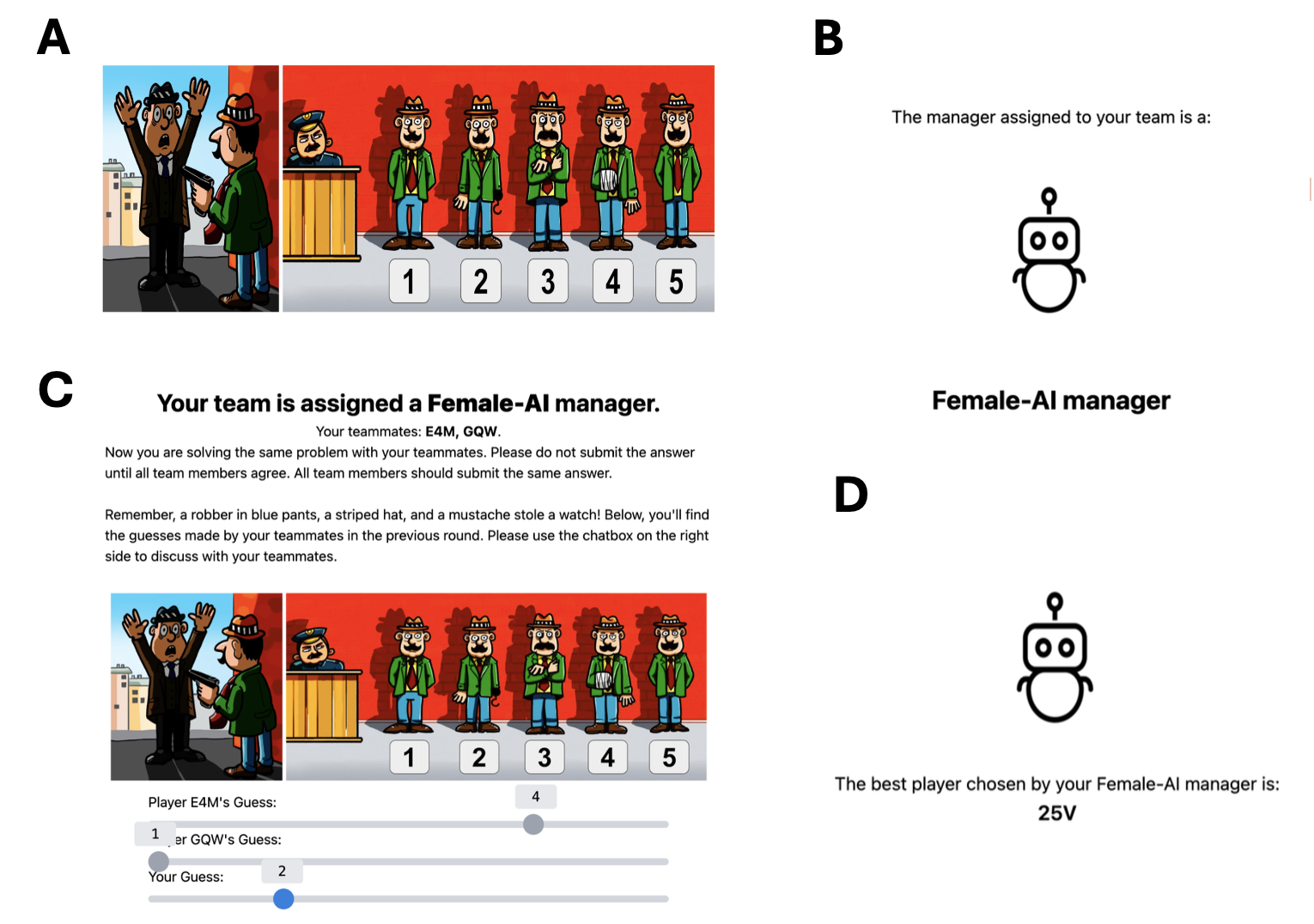} 
	\caption{\textcolor{black}{Key partial views of the experiment interface. \textbf{(A)} Image of the individual task. \textbf{(B)} Manager assignment. \textbf{(C)} Team task. \textbf{(D)} Winner announcement.}}
	\label{fig:task} 
\end{figure}

\textcolor{black}{The experiment consisted of an entry survey, two rounds of the ``Find the Robber" task, and an exit survey. The task was adapted from a publicly available visual reasoning puzzle published on Bright Side~\cite{BrightSide2018}. The images and accompanying scenarios were modified to fit the experimental format and to simulate collaborative problem-solving and managerial evaluation. In the first round (individual round), each participant viewed an image accompanied by a short scenario and was asked to identify ``the robber” from five cartoon suspects by adjusting a slider bar. They were required to submit the answer within a 2-minute limit. In the second round (team round), participants worked in teams of three and were asked to complete the same task collaboratively. A text-based chat interface allowed teammates to discuss their reasoning and jointly select a final team answer within a 4-minute limit. Figure. \ref{fig:task} shows key partial views of the experiment interface (see SI for complete screenshots and further information).}

\textcolor{black}{While the ``Find the Robber" task is relatively simple and does not require advanced mathematics or logic, it engages key cognitive and social processes: perceptual discrimination, judgment under uncertainty (no obvious or guaranteed correct answer), time pressure, and social inference (how participants justify their reasoning and assess others' input). Despite its simplicity compared with real-world workplace tasks, this task provides a controlled and interpretable model for studying reactions to award decisions in ambiguous, team-based settings where credit attribution is inherently uncertain.
}
\subsection{Procedure} 


\textcolor{black}{As shown in Fig. \ref{fig:fig_game}, the experiment followed a structured sequence of stages designed to simulate a team-based decision-making scenario involving managerial evaluation.}

\noindent\textbf{Participant onboarding and pre-treatment survey.}
The experiment took place entirely online and was designed to simulate a decision-making scenario involving teamwork and managerial evaluation. 
Participants recruited via Prolific began by clicking a provided link to enter the Empirica platform \textcolor{black}{(SI Fig. S1)}, where the experiment was hosted.
Upon joining, participants first read an information sheet outlining the study's goals and procedures, followed by providing their consent \textcolor{black}{(SI Figs. S2-S3)}. \textcolor{black}{After entering their Prolific ID, they proceed to the experiment (SI Figs. S4-S5).}
They then completed a brief survey collecting demographic information, such as age, gender, education level, and their initial perceptions of trustworthiness, competence, fairness, and willingness to work with both AI and human managers, as well as male and female managers. This initial survey helped establish baseline perspectives toward different manager types. 
\vspace{0.3cm}

\noindent\textbf{Individual problem-solving task.}
After completing the preliminary survey, participants began the first round of the game.
\textcolor{black}{Each participant saw a countdown timer appeared at the top center of their screen, their randomly assigned ID, a short text description, a cartoon picture, a slider tool, and a submit button (SI Fig. S6).} They were required to individually complete the visual intelligence task (``Find the Robber" ): read the on-screen description, examine the picture based on visual cues, and identify the robber from five options. Participants selected the suspect they believed to be the robber using the slider bar within a 2-minute time limit; otherwise, a default value was automatically recorded.
\vspace{0.3cm}

\noindent\textbf{Manager assignment.}
After completing the \textcolor{black}{first round of the game}, participants were informed they would join a team with two teammates for the next round (SI Figs S7--S10). Each team was randomly assigned a manager, either a human or an AI, who would evaluate the team's performance. The manager's gender was also manipulated within each manager type, with the manager identified as male, female, or gender unspecified. The manager would ultimately select the ``best player" in the team to receive an extra cash award of \pounds0.5.
\vspace{0.3cm}

\noindent\textbf{Team-based task and collaboration.} \textcolor{black}{Following the manager assignment, participants proceeded to the second round of the game.}
In the second round, participants collaborated in a team-based setting through an online chat interface to complete the same task as in the first round (SI Fig. S11). They could view their own and their teammates' responses from the previous round on the slider bars, providing context for discussion. Through the chat, participants worked collectively to arrive at a final team response, simulating real-world team dynamics. All teams completed identical tasks, ensuring a consistent basis for comparing perceptions of the managers across groups.

\vspace{0.3cm}

\noindent\textbf{Contribution assessment.}
After completing the team task, participants rated their own and their teammates' contributions on a scale from 0 to 10 (SI Fig. S12). The chat history remained visible during this rating phase but was disabled for further input, ensuring that participants could not influence each other's ratings. This contribution assessment allowed for an additional evaluation of how individuals perceived their own and their peers' contributions within the team.
\vspace{0.3cm}

\noindent\textbf{Manager decision and exit survey.}
After a waiting time of 20 seconds, participants received the announcements of the ``best player" selected by the manager (SI Figs. S13--S14).  
Finally, participants completed an exit survey that included the study's primary measures: perceptions of trustworthiness, competence, and fairness regarding the manager's decision, as well as their willingness to work with a similar manager in the future. 
\vspace{0.3cm}

\noindent\textbf{Debriefing and payment.} 
After completing the survey, participants were informed during the debriefing that no actual manager was involved in the experiment and that the selection was random. The entire experiment took approximately 12 minutes, providing an experience of teamwork, evaluation, and managerial decision-making. Participants were compensated at a rate of \pounds9 per hour, prorated for the duration of the study. The chosen ``best player" received an additional \pounds0.5 bonus.

\section{Results}

\subsection{Comparison of pre- and post-treatment perceptions}

\begin{figure}[htbp] 
	\centering
	\includegraphics[width=\textwidth]{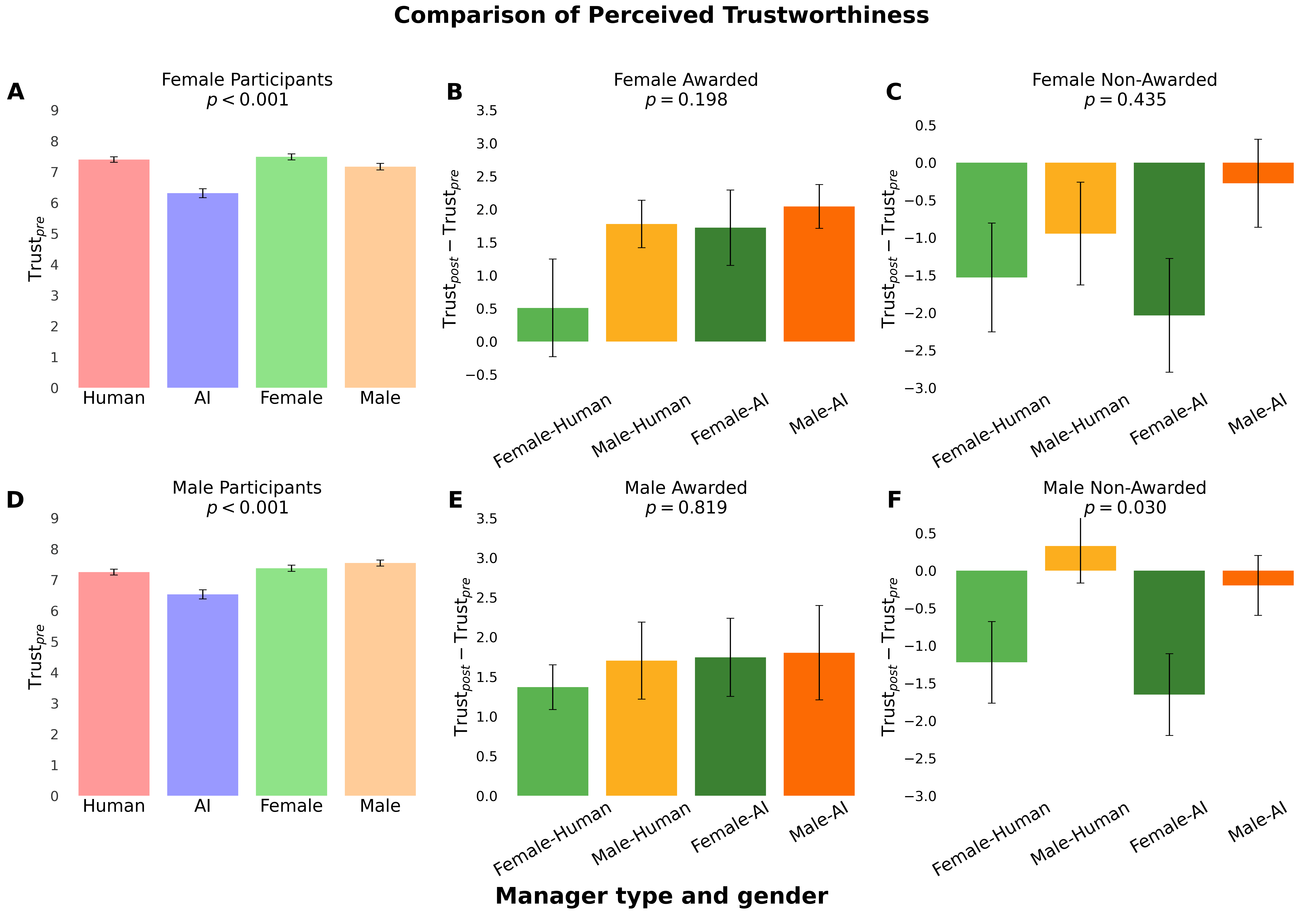} 
	\caption{Comparison of the mean perceived trustworthiness of different groups by manager type and gender. 
		The upper row represents results for female participants, and the lower row represents results for male participants. (A) and (D) show the mean pre-treatment perceived trustworthiness. (B) and (E) depict the mean change in post-treatment perceived trustworthiness for awarded participants, while (C) and (F) illustrate the change for non-awarded participants. 
		Error bars represent the standard error of the mean. 
	} 
	\label{fig:trust} 
\end{figure} 

\begin{figure}[htbp] 
	\centering
	\includegraphics[width=\textwidth]{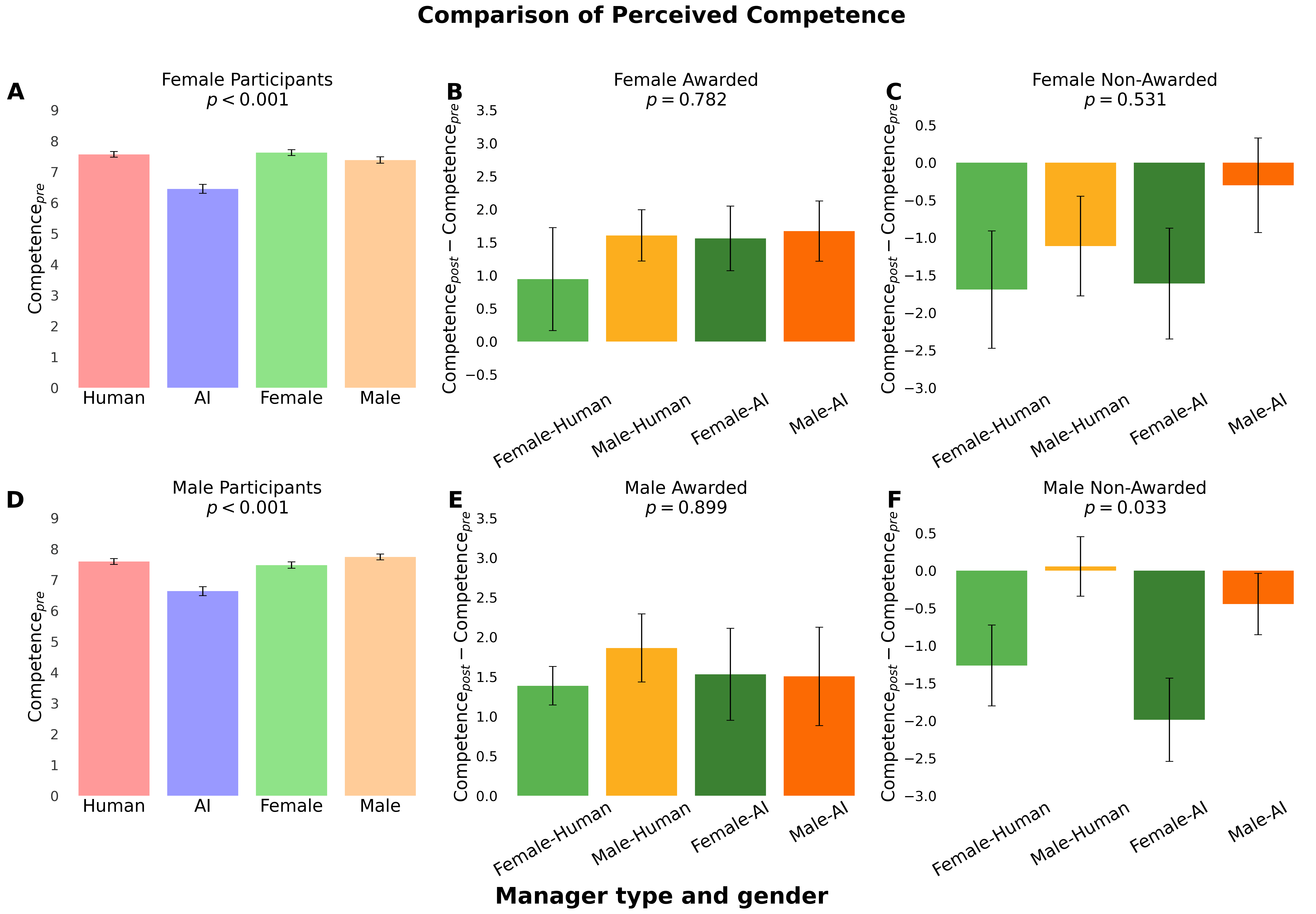} 
	\caption{Comparison of the mean perceived competence of different groups by manager type and gender. 
		The upper row represents results for female participants, and the lower row represents results for male participants. (A) and (D) show the mean pre-treatment perceived competence. (B) and (E) depict the mean change in post-treatment perceived competence for awarded participants, while (C) and (F) illustrate the change for non-awarded participants.
		Error bars represent the standard error of the mean. 
	} 
	\label{fig:competence} 
\end{figure} 

\begin{figure}[htbp] 
	\centering
	\includegraphics[width=\textwidth]{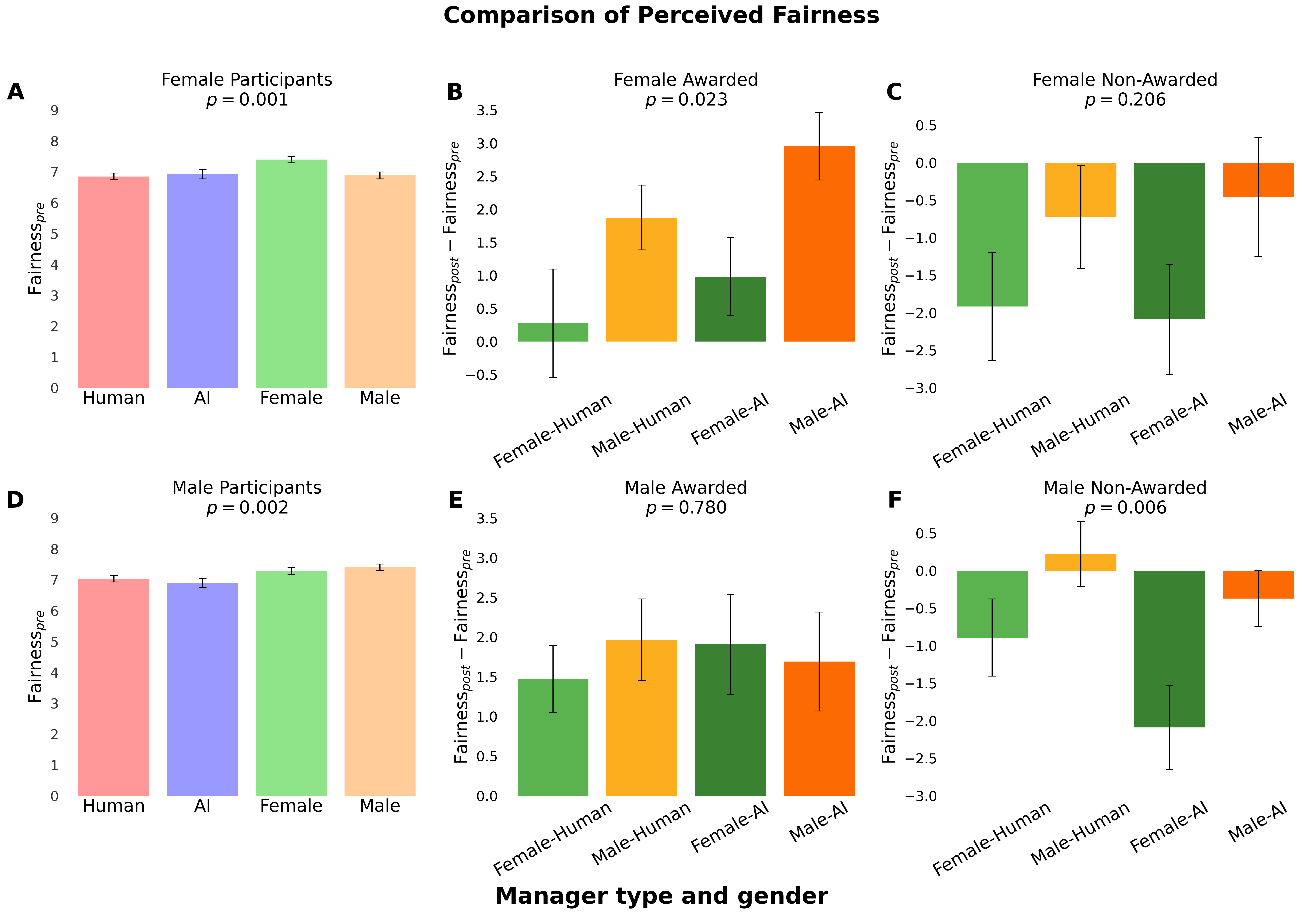} 
	\caption{Comparison of the mean perceived fairness of different groups by manager type and gender. 
		The upper row represents results for female participants, and the lower row represents results for male participants. (A) and (D) show the mean pre-treatment perceived fairness. (B) and (E) depict the mean change in post-treatment perceived fairness for awarded participants, while (C) and (F) illustrate the change for non-awarded participants.
		Error bars represent the standard error of the mean. 
	} 
	\label{fig:fairness} 
\end{figure}

\begin{figure}[htbp] 
	\centering
	\includegraphics[width=\textwidth]{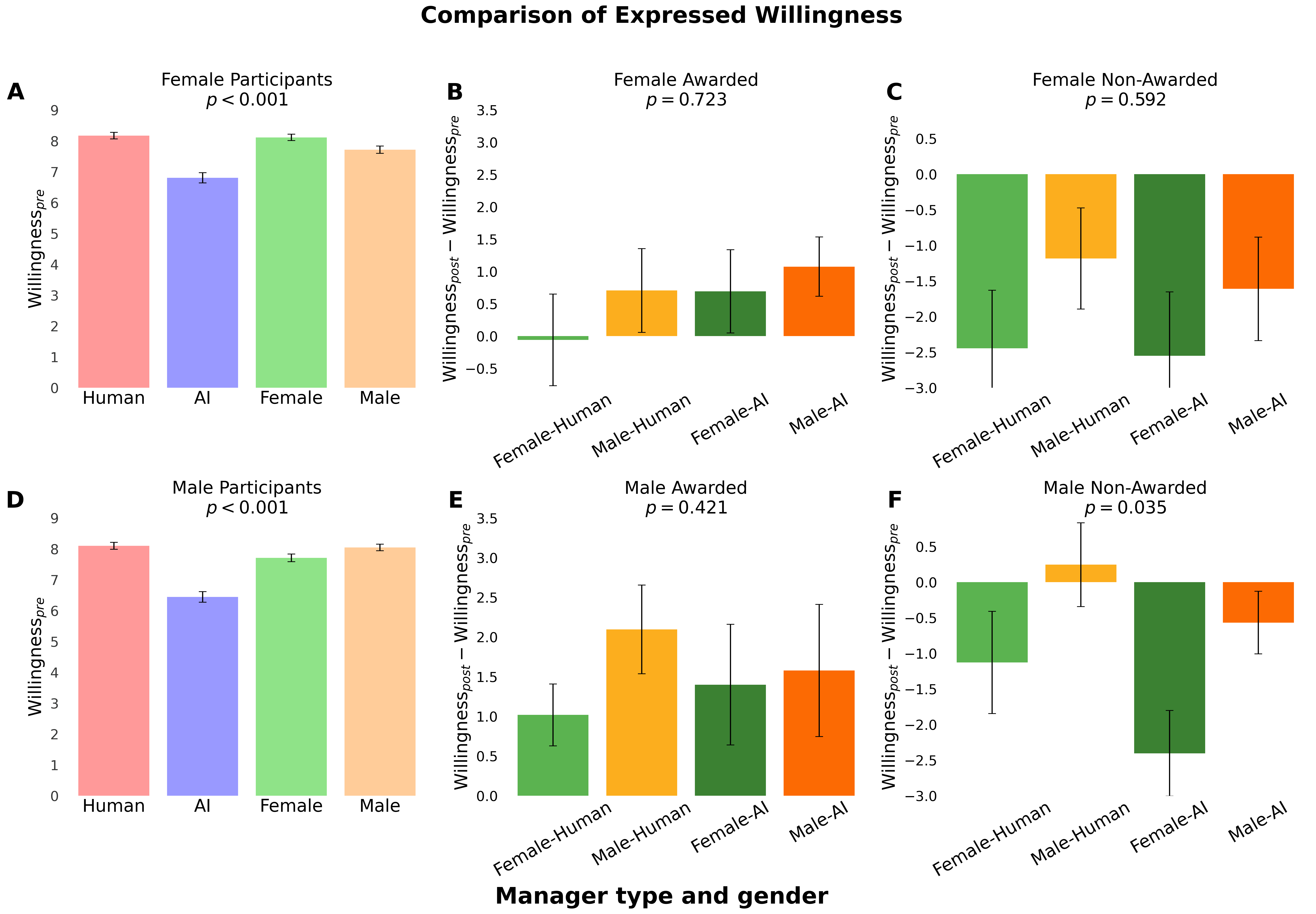} 
	\caption{Comparison of the mean expressed willingness to work with a similar manager in the future, by manager type and gender. 
		The upper row represents results for female participants, and the lower row represents results for male participants. (A) and (D) show the mean pre-treatment perceived willingness. (B) and (E) depict the mean change in post-treatment perceived willingness for awarded participants, while (C) and (F) illustrate the change for non-awarded participants.
		Error bars represent the standard error of the mean. 
	} 
	\label{fig:willingness} 
\end{figure}

In the pre-treatment survey, the mean of participants' perceptions across manager types and genders was around 7, with slight variations (Fig.~\ref{fig:trust}--\ref{fig:willingness} A, D). Participants generally rated human managers more favorably than AI managers. Female participants gave slightly higher ratings to female managers, and male participants gave slightly higher ratings to male managers, though the overall gender differences were relatively small.
The perceptions of managerial decisions show a weak positive correlation between human and AI managers but a strong positive correlation between female and male managers (See SI Fig. S16). This suggests that managerial perceptions in the pre-treatment phase may be more influenced by the manager's type than by the manager's gender.

Post-treatment comparisons (Fig.~\ref{fig:trust}--\ref{fig:willingness} B, E) show that receiving an award increased perceptions across all manager types. Male managers are consistently rated higher across all dimensions among both female and male participants. Female participants gave the highest ratings for male-AI, especially in fairness perception. Male participants rated male-human managers more favorably, especially in their willingness to work with them again in the future. Female human managers are consistently rated the lowest across all dimensions among both female and male participants.

Conversely, not receiving an award (Fig.~\ref{fig:trust}--\ref{fig:willingness} C, F) generally reduced perceptions of managers, with the steepest declines observed for female-AI managers across both female and male participants, and a statistically significant effect among male participants \textcolor{black}{(one-way ANOVA, $p < 0.05$)} on all dimensions. 
The dynamic shifts observed in evaluations before and after the treatment emphasize the influence of manager type, gender, and award outcomes.

\subsection{Effects of demographics and pre-treatment perceptions}

\begin{figure}[htbp] 
	\centering
	\includegraphics[width=0.9\textwidth]{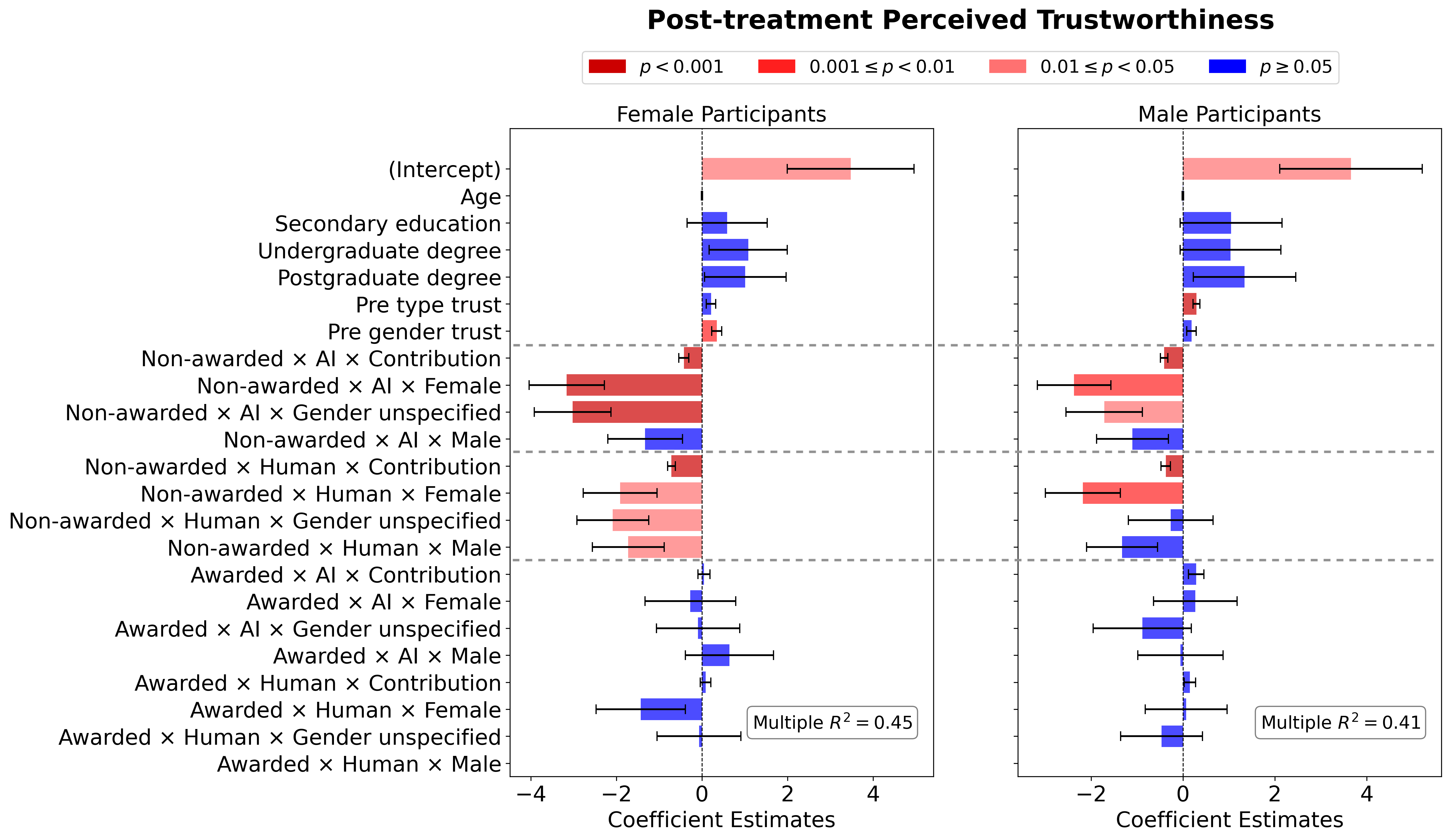} 
	\caption{
		Coefficient estimates ($\beta$) for post-treatment perceived trustworthiness $Trust_{post}$ among female and male participants, based on pre-treatment perceptions, demographic factors, and interactions. 
		Interaction terms capture the influence of manager type, manager gender, and participants' self-perceived contributions on perceived trustworthiness. Statistical significance is denoted by $p<0.05$. Error bars represent standard errors. 
		Dashed lines visually separate pre-treatment factors, non-awarded AI categories, non-awarded human categories, and awarded categories.}
	\label{fig: lm_trust} 
\end{figure} 

\begin{figure}[htbp] 
	\centering
	\includegraphics[width=0.9\textwidth]{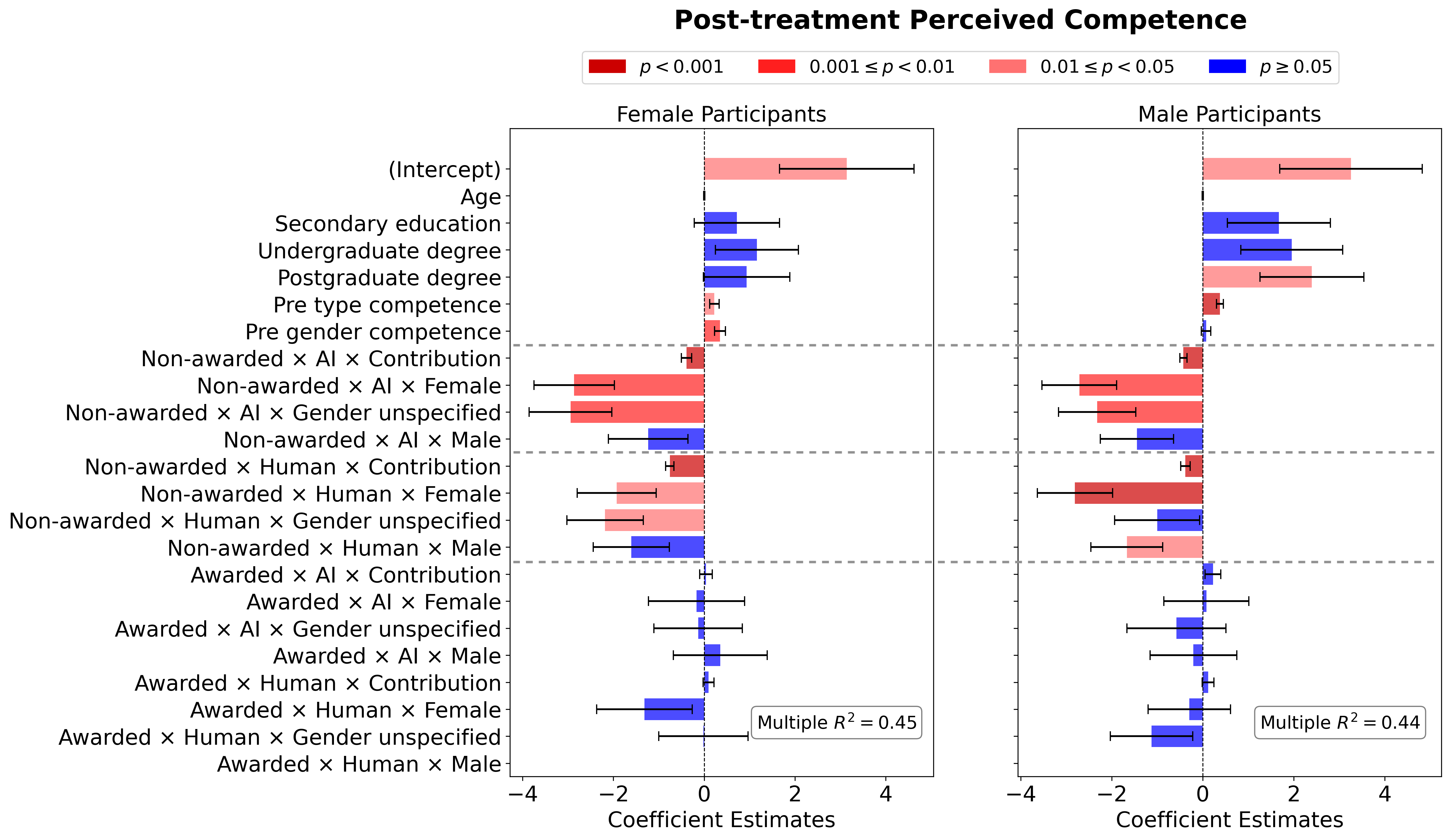} 
	\caption{
		Coefficient estimates ($\beta$) for post-treatment perceived competence $Competence_{post}$ among female and male participants, based on pre-treatment perceptions, demographic factors, and interactions. 
		Interaction terms capture the influence of manager type, manager gender, and participants' self-perceived contributions on perceived competence. Statistical significance is denoted by $p<0.05$. Error bars represent standard errors. 
		Dashed lines visually separate pre-treatment factors, non-awarded AI categories, non-awarded human categories, and awarded categories.}
	\label{fig: lm_competence} 
\end{figure} 

\begin{figure}[htbp] 
	\centering
	\includegraphics[width=0.9\textwidth]{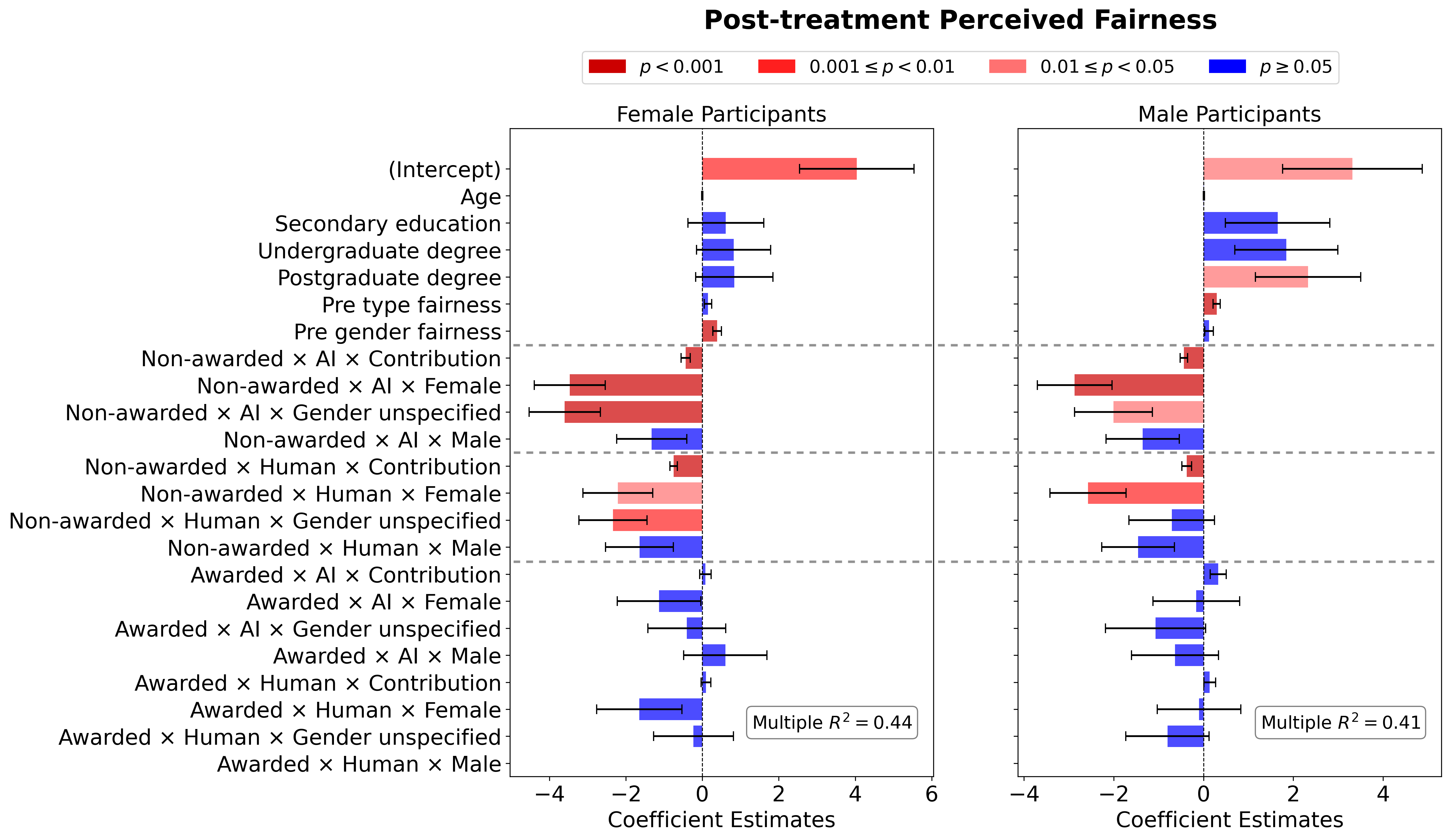} 
	\caption{
		Coefficient estimates ($\beta$) for post-treatment perceived fairness $Fairness_{post}$ among female and male participants, based on pre-treatment perceptions, demographic factors, and interactions.
		Interaction terms capture the influence of manager type, manager gender, and participants' self-perceived contributions on perceived fairness. Statistical significance is denoted by $p<0.05$. Error bars represent standard errors.
		Dashed lines visually separate pre-treatment factors, non-awarded AI categories, non-awarded human categories, and awarded categories.}
	\label{fig:fig2} 
\end{figure} 

\begin{figure}[htbp] 
	\centering
	\includegraphics[width=0.9\textwidth]{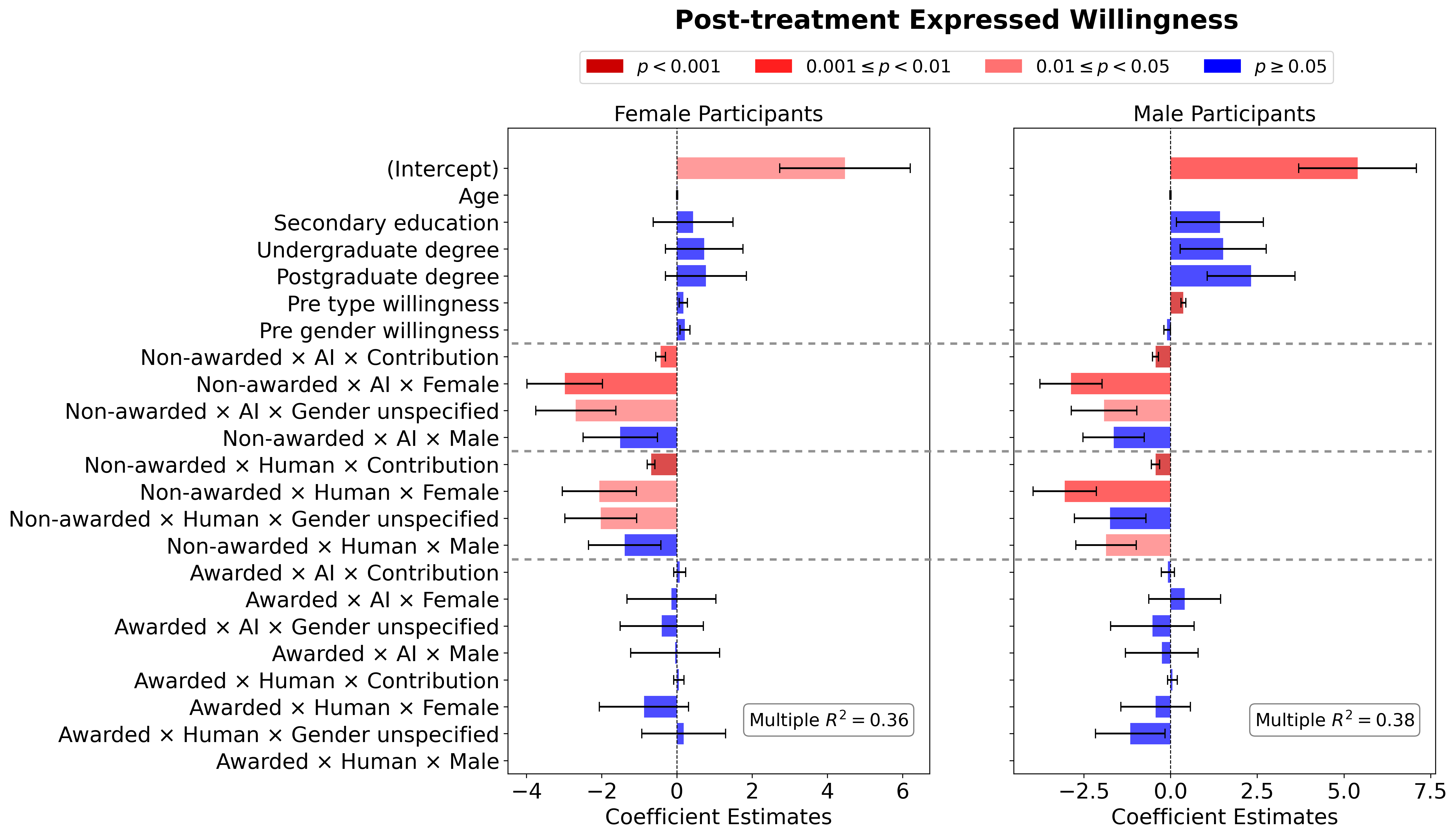} 
	\caption{
		Coefficient estimates ($\beta$) for post-treatment expressed willingness $Willingness_{post}$ among female and male participants, based on pre-treatment perceptions, demographic factors, and interactions. 
		Interaction terms capture the influence of manager type, manager gender, and participants' self-perceived contributions on expressed willingness. Statistical significance is denoted by $p<0.05$. Error bars represent standard errors. 
		Dashed lines visually separate pre-treatment factors, non-awarded AI categories, non-awarded human categories, and awarded categories.}
	\label{fig: lm_willingness} 
\end{figure}

Figures \ref{fig: lm_trust}--\ref{fig: lm_willingness} illustrate the factors influencing post-treatment perceptions according to the linear regression model with interaction terms (Eq.~\ref{eq:post_fairness}) for female and male participants separately. Post-treatment perceptions are shaped by different persistent factors. For female participants, pre-treatment perceptions of the manager's gender-related trustworthiness,  competence, and fairness significantly predict the corresponding post-treatment perceptions. In contrast, for male participants, pre-treatment perceptions of the manager's type play a stronger role in shaping the post-treatment perceptions of all four dimensions. Demographic variables such as age and education generally show no significant effect on participants' perceptions of their manager's decision, except for male participants, where a postgraduate degree is associated with higher post-treatment ratings of managerial competence and fairness.

\subsection{Interaction effects: Award Outcome $\times$ Manager Type $\times$ Manager Gender}

As shown in Figs. \ref{fig: lm_trust}--\ref{fig: lm_willingness}, relative to the reference category (awarded by a male-human manager), not receiving an award significantly lowers perceptions across all dimensions for female-AI managers, female-human managers, and AI managers of unspecified gender, for both female and male participants. Interestingly, for female participants, not receiving an award also significantly lowers perceptions of all dimensions for human managers of unspecified gender. 
When not awarded by a male-human manager, female participants showed a significant decline in trust, while male participants rated the manager as less competent and expressed lower willingness to work with a similar manager in the future. Detailed statistics are shown in the Supplementary Information Tabs. S1-S8.

\subsection{Interaction effects: Award Outcome $\times$ Manager Type $\times$ Contribution}

Interaction effects between participants' self-perceived contribution, award outcome, and manager type reveal that 
perceptions towards managers are closely tied to an expectation-reward dynamic. As shown in Figs. \ref{fig: lm_trust}--\ref{fig: lm_willingness}, not receiving an award, under high self-perceived contribution, significantly reduces perceptions on all four dimensions for both human managers and AI managers, and for both female and male participants. 

The consistent reduction in the evaluations of human managers among female participants is greater than that for AI managers. It exceeds the reduction observed for either human or AI managers among male participants.
This may stem from higher expectations 
in human leadership from female participants. See Supplementary Information Tabs. S1-S8 for detailed statistics.

\subsection{\textcolor{black}{Participant gender differences}}

\textcolor{black}{Female participants and male participants showed comparable patterns in managerial perceptions, though differences also emerged. Female participants who perceived themselves as high contributors rated human managers more negatively when not awarded. 
	Male participants, when not awarded by a male human manager, rated that manager as significantly less competent and reported lower willingness to work with similar managers in the future. Additionally, postgraduate education among male participants predicted more favorable post-treatment ratings of managerial competence and fairness.}

\section{Discussion}

\textcolor{black}{The findings show that gender bias in leadership evaluations extends beyond human managers to AI decision-makers. 
	Receiving an award increased participants' positive perceptions of the manager across all manager types, whereas not receiving an award led to lower evaluations. However, the magnitude of these changes was moderated by manager gender. Male managers, whether human or AI, benefited more from positive outcomes and were less penalized for negative ones than female managers. The results provide new insights into how social stereotypes and expectations shape human perceptions of AI in leadership roles.}

\subsection{\textcolor{black}{Theoretical and organizational implications}} 

Male managers exhibited greater resilience to negative outcomes, likely due to societal leniency afforded to male figures, consistent with status expectations theory~\cite{ridgeway2001gender}. 
The results align with pervasive societal stereotypes that link competence and leadership effectiveness with male figures~\cite{eagly1984gender, ridgeway2001gender, eagly2002role}. The reinforcement of positive recognition for male managers could reflect societal tendencies to perceive male-led successes as more legitimate or expected, a bias that some have argued is deeply ingrained in patriarchal organizational structures~\cite{connell2020social}. 

Negative outcomes notably harmed perceptions of female managers, particularly female-AI managers, highlighting a compounded vulnerability, where skepticism toward AI decision-making intersects with gender biases to yield harsher judgments. Female human managers also faced a notable decline in the assessment of participants, though to a lesser extent, possibly due to the perception that human managers demonstrate greater accountability and empathy~\cite{montemayor2022principle}.
This underscores the moderating role of gender in shaping perceptions of managerial decisions, illustrating a deeply embedded double standard where cultural narratives shield male leaders with presumed competence and an external locus of control. In contrast, female leaders face greater blame for unfavorable outcomes. 
These findings also resonate with broader feminist critiques that women's roles and societal positions are often disproportionately scrutinized within a male-dominated societal framework~\cite{butler1990feminism, hooks2000feminist}.

\textcolor{black}{The results contribute to theory by extending role congruity and equity frameworks into the domain of AI leadership. 
	They show that people’s perceptions of the managers are shaped by both the outcome of the decision and their gender-related expectations -- even when the manager is non-human. This suggests that the same social and psychological biases that influence how we evaluate human managers also shape how we perceive artificial decision-makers.}

\textcolor{black}{For organizations, this has tangible implications. As AI systems take on evaluative and supervisory roles, employee trust and morale will depend on perceived fairness and social legitimacy of these systems. Training programs and workplace policies should therefore address not only algorithmic bias but also user bias -- how workers interpret and respond to AI managers based on explicit or assumed gender cues.}

\textcolor{black}{Beyond leadership and stereotype frameworks, our findings also align with broader literature on gendered experiences with digital technologies and institutional approaches to algorithmic bias. Prior research on digital inequality shows that gender shapes how individuals interact with digital technologies, including differences in online privacy protection behavior~\cite{park2015men}, and how digital platforms reflect or amplify pre-existing biases ~\cite{hargittai2006differences,schellekens2019female}, reflecting broader socialized gender norms in digital contexts. From an institutional perspective, recent work on algorithmic bias and diversity regulation highlights how conceptualizations of diversity influence efforts to mitigate bias in AI systems~\cite{park2024tprc}. Viewed through this lens, the gendered perceptions of AI managers observed in this study reflect not only individual stereotype transfer but also wider sociotechnical patterns in which longstanding social norms become embedded in emerging algorithmic systems \cite{suchman2007human}.}

\subsection{\textcolor{black}{Implications for AI leadership and design}} 

\textcolor{black}{The findings show that AI managers inherit the same gendered expectations that shape human leadership judgments. The amplified negative impact on female-AI managers emphasizes the need to address intersectional biases in workplace dynamics, where gender and technology intersect to create compounded disadvantages. For designers and organizations, this underscores the importance of bias-aware AI interface design. Perceptions of AI managers depend not only on their visual or anthropomorphic cues, but also on how users interpret these features through the lens of human managerial norms and social expectations. Organizations aiming to integrate AI into leadership roles should consider strategies to mitigate such biases, such as exploring gender-neutral visual and verbal cues and fostering a culture that values fairness and equity. }

\subsection{\textcolor{black}{Limitations}}

\textcolor{black}{Several limitations of the current study should be acknowledged. First, the study relied on an online participant pool (Prolific) composed primarily of digitally literate U.S. adults. Such samples may not fully represent broader workforce demographics, potentially biasing perceptions toward more tech-familiar or culturally homogeneous views.}
Second, our findings, which were based on an experiment conducted in a single setting with a specific task, may not generalize to more complex scenarios involving diverse tasks, larger groups, or mixed-agent interactions between humans and AI in the same environment. 
\textcolor{black}{Third, participants were aware they were in a simulated experiment, which may limit ecological validity. Field studies or hybrid human-AI team experiments in actual workplaces could enhance realism.}
\textcolor{black}{Finally, as with many online studies, we cannot fully exclude the possibility that participants used generative AI tools (e.g., ChatGPT or Gemini). However, the nature of the task, requiring visual inspection, individual judgment, and real-time discussion, makes such use unlikely to have influenced the results.}

\subsection{\textcolor{black}{Future directions}}

Future work could expand on our design by exploring tasks across a broader spectrum of complexity, such as logic-based puzzles, creative group challenges, or multi-stage decision processes that require more sustained effort, information sharing, and interdependence among group members. This would help determine whether the observed patterns in perception and bias generalize to more complex and ecologically valid settings.

Further research could investigate the mechanisms underlying the observed biases, such as cognitive biases, stereotypes, and trust dynamics, to gain a deeper understanding of how they influence various dimensions of perception, including trustworthiness, competence, fairness, and willingness to collaborate. Broader contextual factors, including cultural norms, team dynamics, and prior experiences with AI or human managers could also be examined to uncover the nuanced drivers of these perceptions. 
\textcolor{black}{Future studies could employ cross-cultural samples and longitudinal designs to examine how these biases evolve and to test interventions such as anonymized AI presentation or variations in AI competence framing as potential strategies for reducing bias.}
Insights from such work will be essential for designing systems across a range of settings that foster equitable, trustworthy, unbiased, and effective integration of AI in managerial and collaborative tasks.

The emergence of AI in managerial and societal roles presents an opportunity to move beyond outdated frameworks and mindsets, yet it also risks perpetuating or even amplifying existing gender and sexist stereotypes if left unchecked. While significant scrutiny is placed on ensuring AI is human-centric, fair, and explainable, we must also hold ourselves accountable as humans to embody these principles in our behaviors and societal norms. History has shown how prejudice and discrimination linger in our actions and institutions, and these biases could be magnified as AI becomes more embedded in daily life. To prevent this, it is imperative to address and dismantle these stereotypes now -- before they shape the interactions between humans and AI in ways that are counterproductive or harmful. Policymakers currently shaping AI regulation must consider these risks and explore bold measures, such as developing AI systems that are gender-neutral, to break free from reinforcing stereotypes. By addressing these issues today, we lay the groundwork for a more equitable future where both humans and AI contribute to a society free from discrimination.

\section{Conclusion}

Our study reveals that award outcomes influence perceptions of managers, and these effects are significantly moderated by both the manager's labeled gender and whether the manager is human or AI. Male managers benefited more from positive outcomes, while female managers, particularly female-AI, were disproportionately penalized after unfavorable outcomes. By integrating insights from social psychology, gender studies, and human-AI interaction, our work shows how longstanding societal biases can manifest in emerging hybrid work environments. It highlights the nuanced ways that gender and technology intersect in shaping perceptions of managerial trust, competence, fairness, and future collaboration.

\textcolor{black}{Beyond documenting these patterns, the study contributes to a deeper understanding of how human social cognition transfers to AI contexts, showing that technological neutrality does not eliminate deep-seated human bias.}
As AI systems, and particularly ``AI Agents", take on increasingly prominent roles in decision-making and leadership, addressing these compounded biases is essential. Building equitable, trustworthy, and inclusive systems will require not only fair algorithms but also greater awareness of how human perceptions, shaped by gender norms and societal bias, interact with algorithmic decision-makers and influence how those systems are interpreted and received.

\textcolor{black}{Moving forward, these insights call for a dual approach: designing AI systems that minimize gendered cues, while also fostering workplace cultures that challenge implicit bias and support equitable human-human, and human–AI collaboration. Future research should continue to explore strategies that reduce bias at both the human and system levels, ensuring that the integration of AI in leadership promotes equity rather than reproducing inequality. By addressing these challenges proactively, organizations can move toward a future where both human and AI-led decisions are evaluated fairly and constructively.}

\section*{Data availability}

Experimental results data have been deposited in OSF (\href{https://osf.io/253wq/}{https://osf.io/253wq/})~\cite{database}. 

\section*{ACKNOWLEDGMENTS}
The research conducted in this publication was funded by the Irish Research Council under grant number IRCLA/2022/3217, ANNETTE (Artificial Intelligence Enhanced Collective Intelligence).
We thank Bahador Bahrami and Nico Mutzner for their valuable discussions and comments. 

\bibliographystyle{unsrt}

\bibliography{references}

\clearpage
\appendix
\clearpage
\setcounter{figure}{0}
\renewcommand{\thefigure}{S\arabic{figure}}
\begin{center}
    \textbf{\LARGE SUPPLEMENTARY INFORMATION}
\end{center}
\bigskip
	\renewcommand{\thefigure}{S\arabic{figure}}  
	\renewcommand{\thetable}{S\arabic{table}}

	\section*{Pre-treatment survey questions}
	
	\begin{enumerate}
		\item Age (Participants can enter a value greater than or equal to 18 and less than or equal to 100)
		\item What is your gender identity?
		\\ (Options: Male, Female, Non-binary, Other)
		
		\item What is the highest level of education you have completed?
		\\ (Options: Secondary education, Undergraduate degree (e.g., BA, BSc), Postgraduate degree (e.g., MA, MSc, PhD), Other)
		
		\item How \textbf{trustworthy} do you think an experienced \textbf{``human"} manager is in making award decisions for the team they manage? 0: very untrustworthy; 10: very trustworthy
		\\ (Options from 0 to 10)
		
		\item How \textbf{trustworthy} do you think a trained \textbf{``AI"} manager is in making award decisions for the team they manage? 0: very untrustworthy; 10: very trustworthy
		\\ (Options from 0 to 10)
		
		\item How \textbf{competent} do you think an experienced \textbf{``human"} manager is in making award decisions for the team they manage? 0: very incompetent; 10: very competent
		\\ (Options from 0 to 10)
		
		\item How \textbf{competent} do you think a trained \textbf{``AI"} manager is in making award decisions for the team they manage? 0: very incompetent; 10: very competent
		\\ (Options from 0 to 10)
		
		\item How \textbf{fair} do you think an experienced \textbf{``human"} manager is in making award decisions for the team they manage? 0: very unfair; 10: very fair
		\\ (Options from 0 to 10)
		
		\item How \textbf{fair} do you think a trained \textbf{``AI"} manager is in making award decisions for the team they manage? 0: very unfair; 10: very fair
		\\ (Options from 0 to 10)
		
		\item Would you be \textbf{willing} to work in a small team led by an experienced \textbf{``human"} manager who makes award decisions for the team they manage? 0: very unwilling; 10: very willing
		\\ (Options from 0 to 10)
		
		\item Would you be \textbf{willing} to work in a small team led by a trained \textbf{``AI"} manager who makes award decisions for the team they manage? 0: very unwilling; 10: very willing
		\\ (Options from 0 to 10)
		
		\item How \textbf{trustworthy} do you think an experienced \textbf{``male"} manager is in making award decisions for the team they manage? 0: very untrustworthy; 10: very trustworthy
		\\ (Options from 0 to 10)
		
		\item How \textbf{trustworthy} do you think an experienced \textbf{``female"} manager is in making award decisions for the team they manage? 0: very untrustworthy; 10: very trustworthy
		\\ (Options from 0 to 10)
		
		\item How \textbf{competent} do you think an experienced \textbf{``male"} manager is in making award decisions for the team they manage? 0: very incompetent; 10: very competent
		\\ (Options from 0 to 10)
		
		\item How \textbf{competent} do you think an experienced \textbf{``female"} manager is in making award decisions for the team they manage? 0: very incompetent; 10: very competent
		\\ (Options from 0 to 10)
		
		\item How \textbf{fair} do you think an experienced \textbf{``male"} manager is in making award decisions for the team they manage? 0: very unfair; 10: very fair
		\\ (Options from 0 to 10)
		
		\item How \textbf{fair} do you think an experienced \textbf{``female"} manager is in making award decisions for the team they manage? 0: very unfair; 10: very fair
		\\ (Options from 0 to 10)
		
		\item Would you be \textbf{willing} to work in a small team led by an experienced \textbf{``male"} manager who makes award decisions for the team they manage? 0: very unwilling; 10: very willing
		\\ (Options from 0 to 10)
		
		\item Would you be \textbf{willing} to work in a small team led by an experienced \textbf{``female"} manager who makes award decisions for the team they manage? 0: very unwilling; 10: very willing
		\\ (Options from 0 to 10)
		
	\end{enumerate}
	
	\section*{Post-treatment survey questions}
	
	\begin{enumerate}
		
		\item How trustworthy do you think the manager was in making award decision for the best player in the game? 0: very untrustworthy; 10: very trustworthy
		\\ (Options from 0 to 10)
		
		\item How competent do you think the manager was in making award decision for the best player in the game? 0: very incompetent; 10: very competent
		\\ (Options from 0 to 10)
		
		\item How fair do you think the manager was in making award decision for the best player in the game? 0: very unfair; 10: very fair
		\\ (Options from 0 to 10)
		
		\item Will you be willing to work in future teams with a manager like the one in this game, who makes award decisions? 0: very unwilling; 10: very willing
		\\ (Options from 0 to 10)
		
		\item How satisfied did you feel about the manager’s award decision in the game? 0: very unsatisfied; 10: very satisfied
		\\ (Options from 0 to 10)
		
		\item What was the manager type in your group?
		\\ (Options: Human, AI, No manager)
		
		\item What was the manager gender in your group? (Please select ``Gender unspecified" if no manager was assigned.)
		\\ (Options: Female, Male, Gender unspecified)
	\end{enumerate}
	
	\clearpage
	
	\section*{Key experiment screenshots}
	
	\begin{figure}[htbp] 
		\centering
		\includegraphics[width=\textwidth]{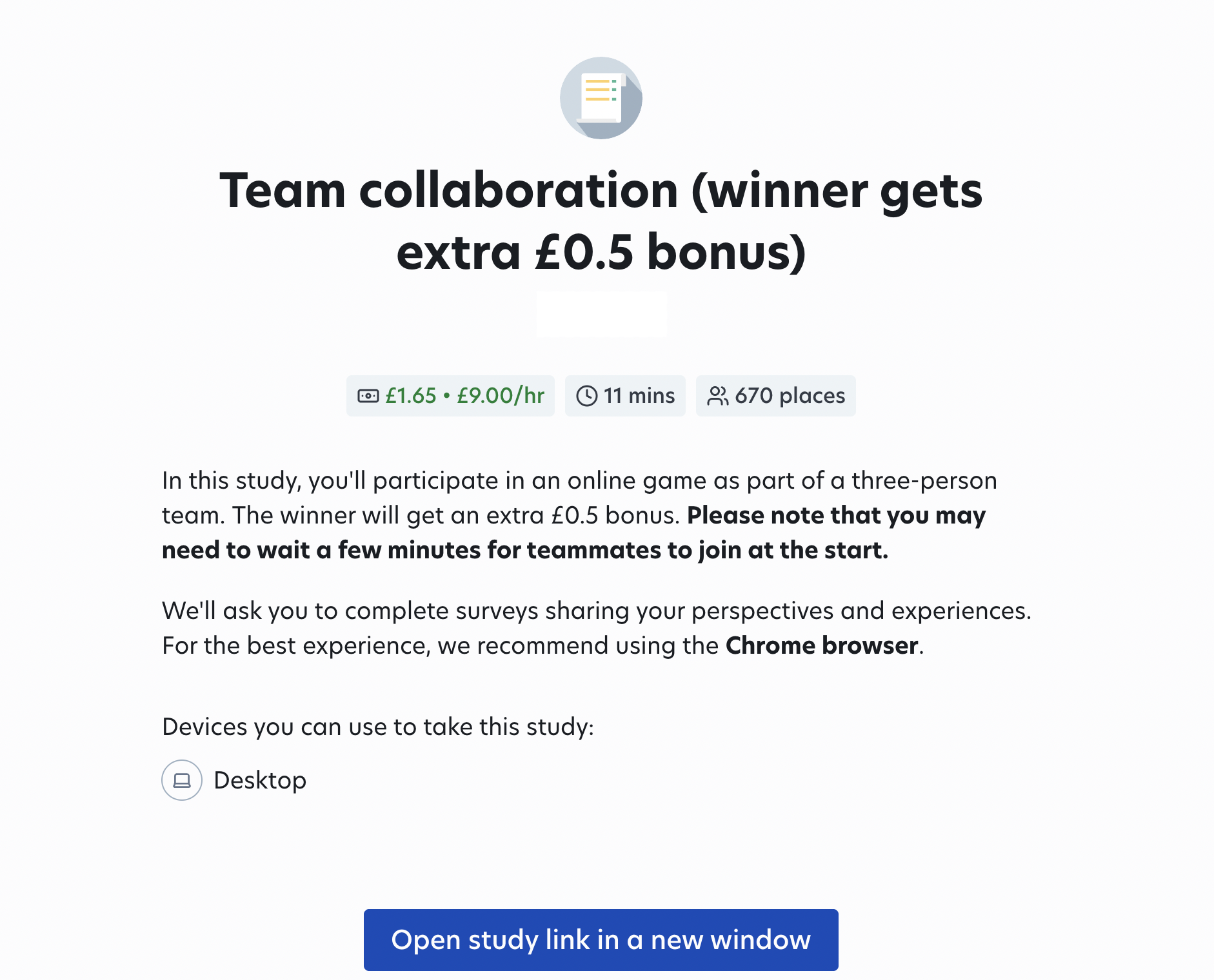} 
		\caption{\textcolor{black}{Participant view on Prolific. Clicking the link directs participants to the information sheet.}} 
		\label{fig:view_prolific} 
	\end{figure}

	\begin{figure}[htbp] 
		\centering
		\includegraphics[width=0.85\textwidth]{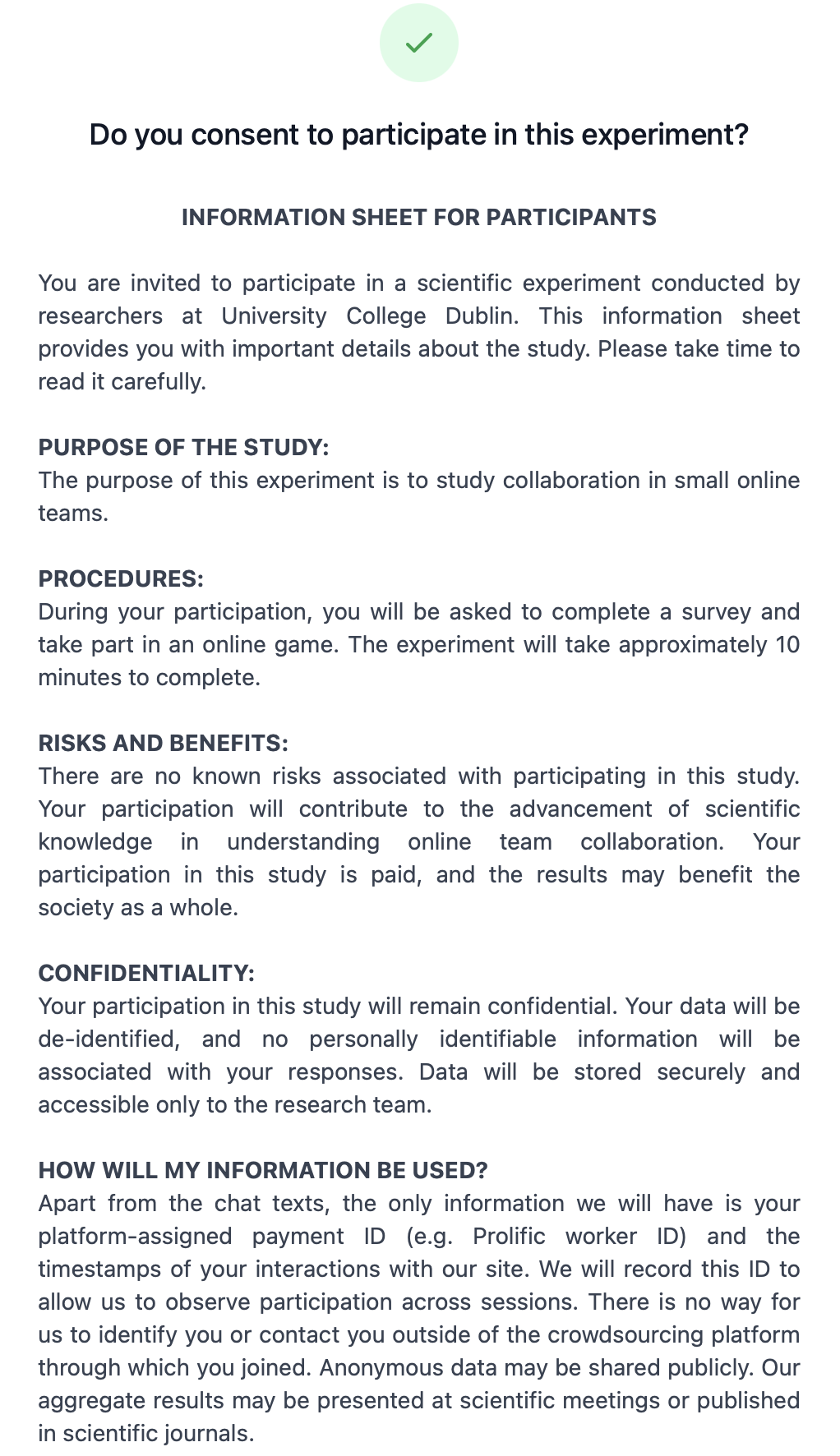} 
		\caption{\textcolor{black}{Information sheet and consent form (part 1).}} 
		\label{fig:consent1} 
	\end{figure} 
	
	\begin{figure}[htbp] 
		\centering
		\includegraphics[width=0.9\textwidth]{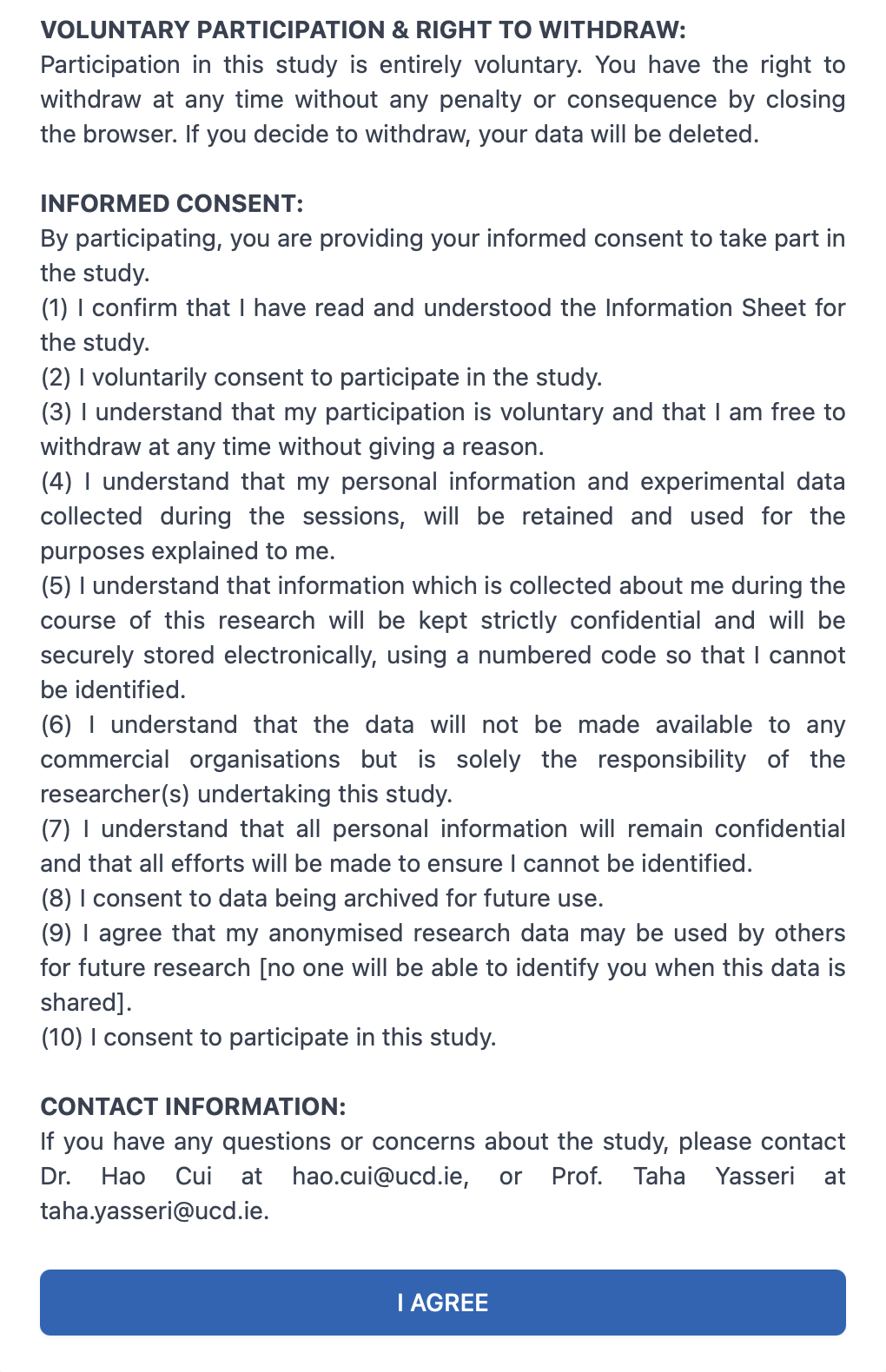} 
		\caption{\textcolor{black}{Information sheet and consent form (part 2). After reading the information, participants click ``I AGREE” to provide consent.}} 
		\label{fig:consent2} 
	\end{figure}

	\begin{figure}[htbp] 
		\centering
		\includegraphics[width=\textwidth]{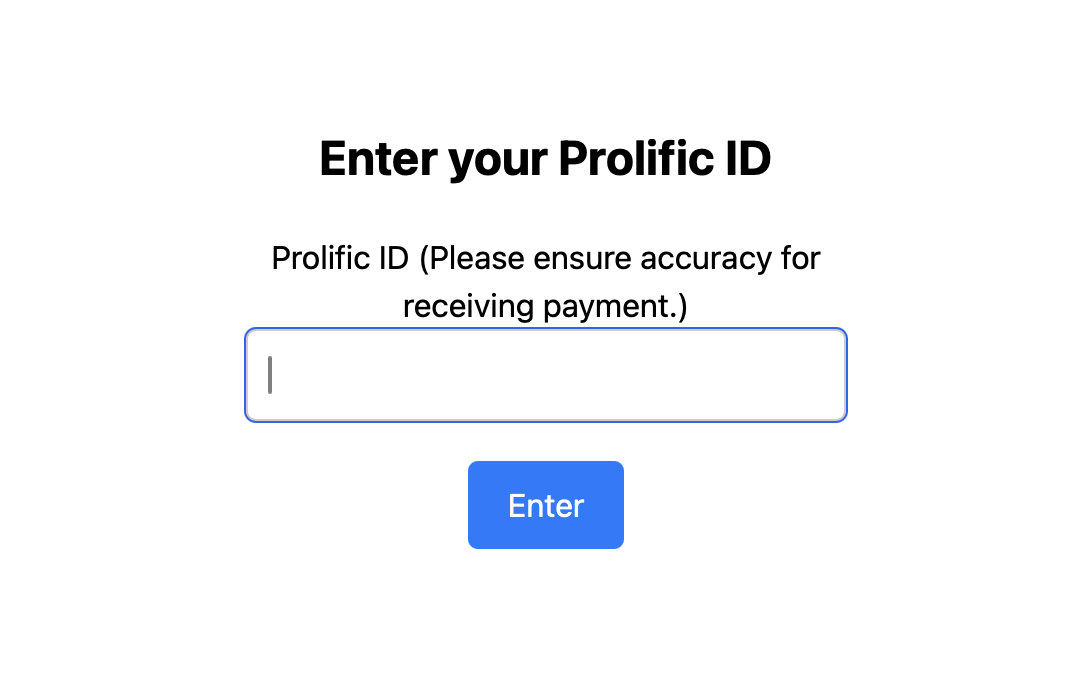} 
		\caption{\textcolor{black}{After providing consent, participants arrive at this page. Upon submitting their Prolific ID, they proceed to the experiment.}} 
		\label{fig:enter_prolific} 
	\end{figure} 
	
	\begin{figure}[htbp] 
		\centering
		\includegraphics[width=\textwidth]{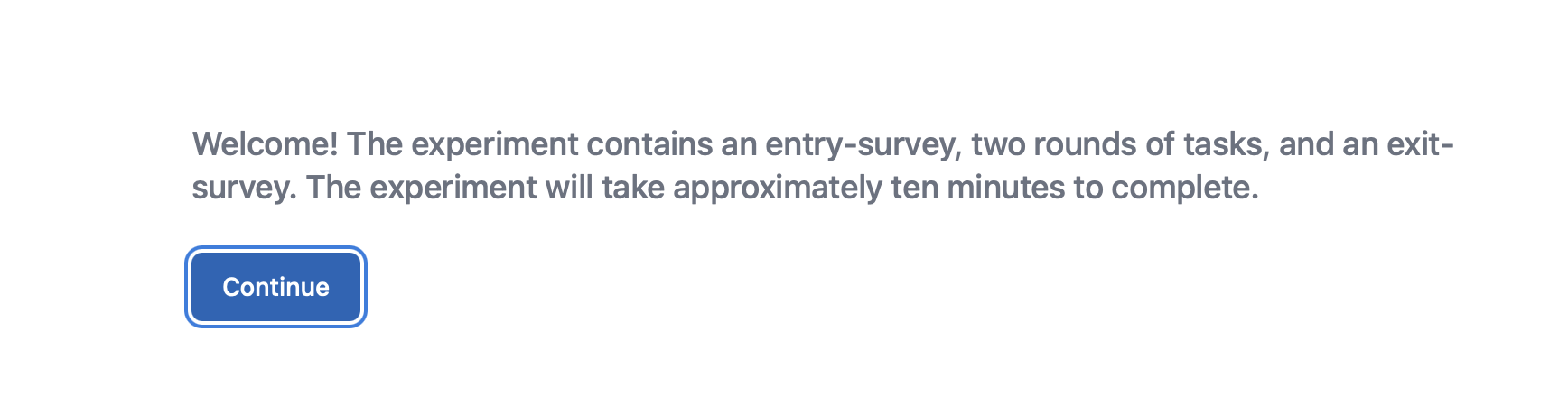} 
		\caption{\textcolor{black}{Welcome page of the experiment. Clicking ``Continue” directs participants to the pre-treatment survey.}} 
		\label{fig:welcome} 
	\end{figure}

	\begin{figure}[htbp] 
		\centering
		\includegraphics[width=\textwidth]{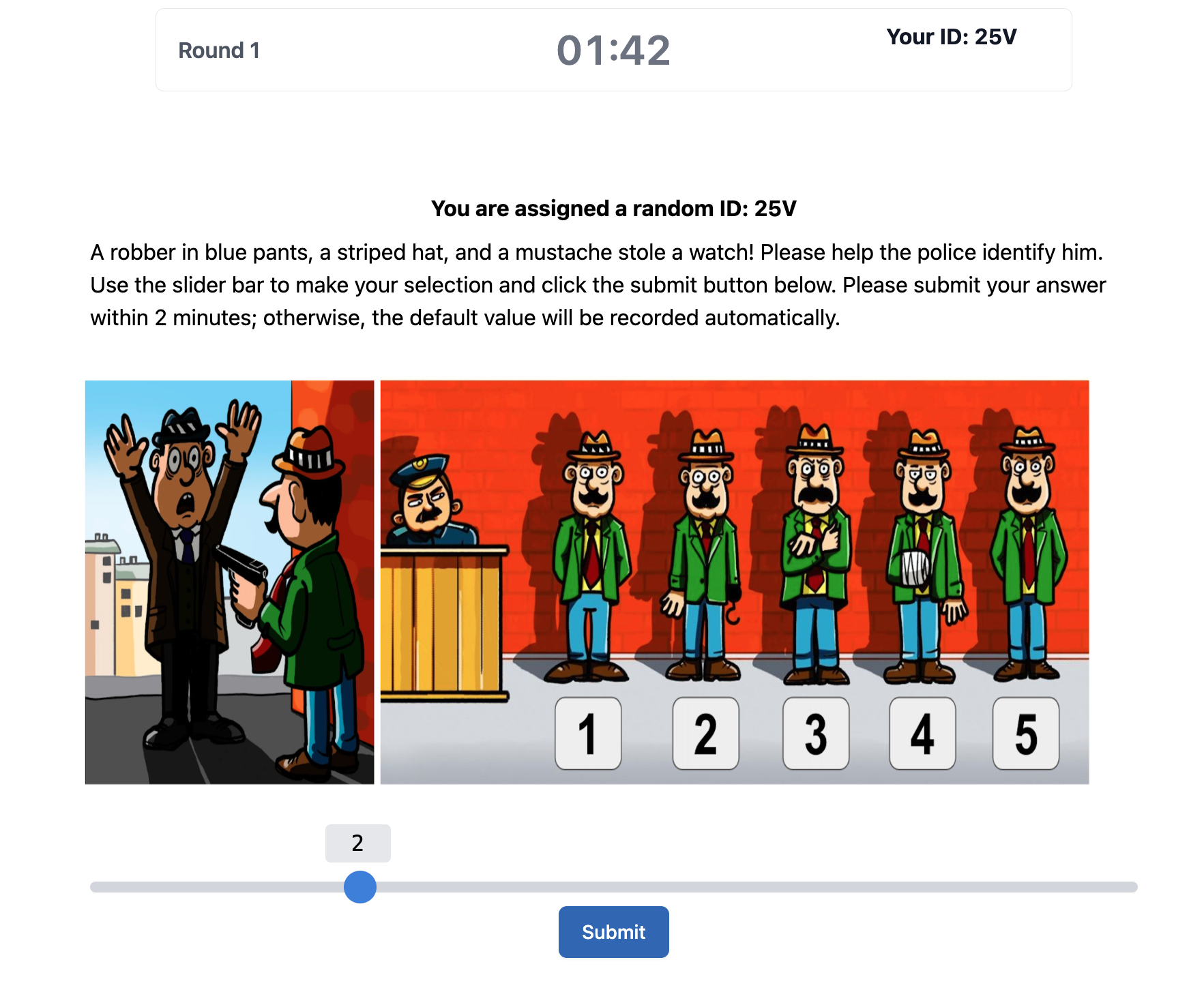} 
		\caption{\textcolor{black}{After completing the pre-treatment survey, participants proceed to the first round of the game.}} 
		\label{fig:page1} 
	\end{figure} 
	
	\begin{figure}[htbp] 
		\centering
		\includegraphics[width=\textwidth]{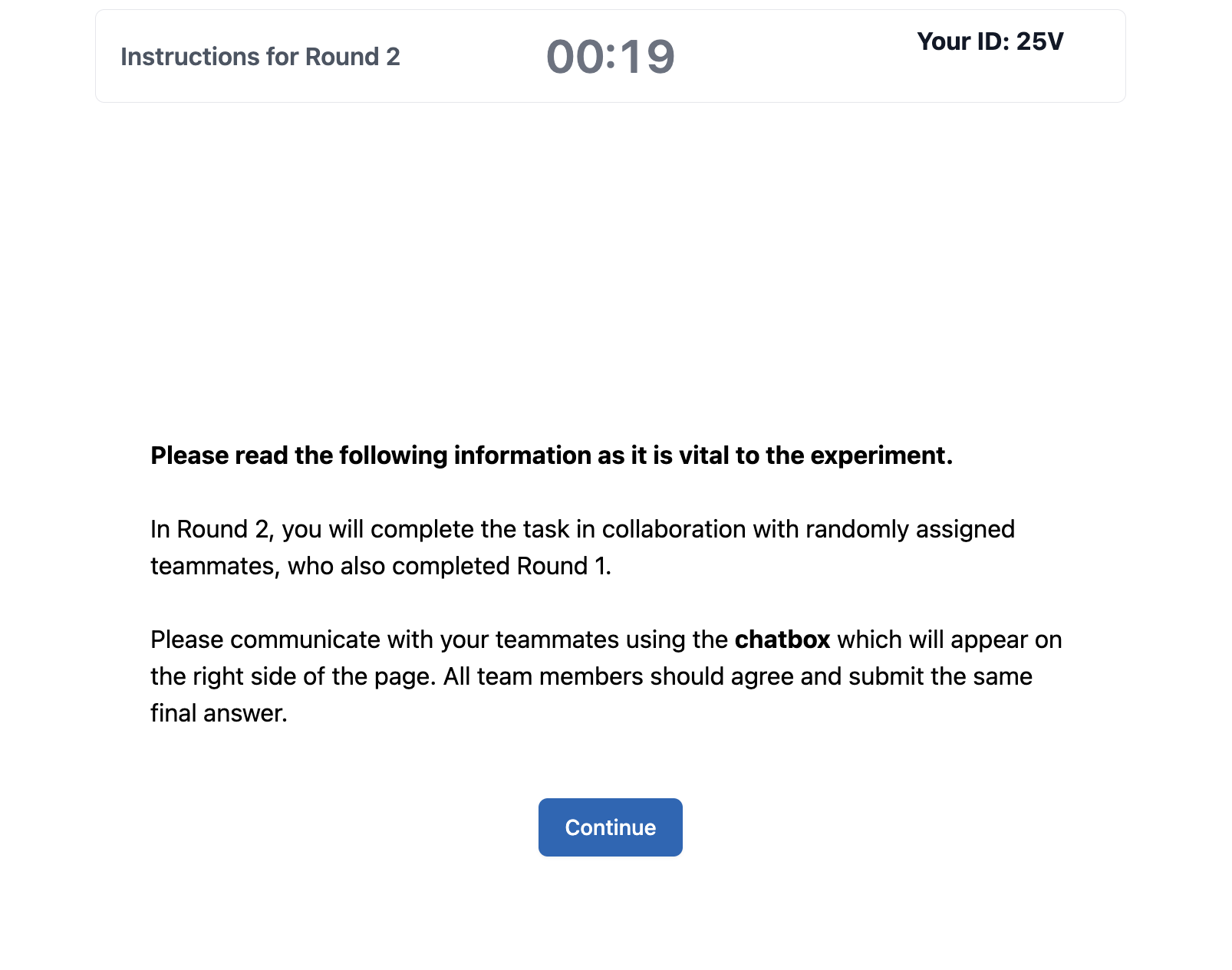} 
		\caption{Instructions for the second round - page 1.} 
		\label{fig:page2} 
	\end{figure} 
	
	\begin{figure}[htbp] 
		\centering
		\includegraphics[width=\textwidth]{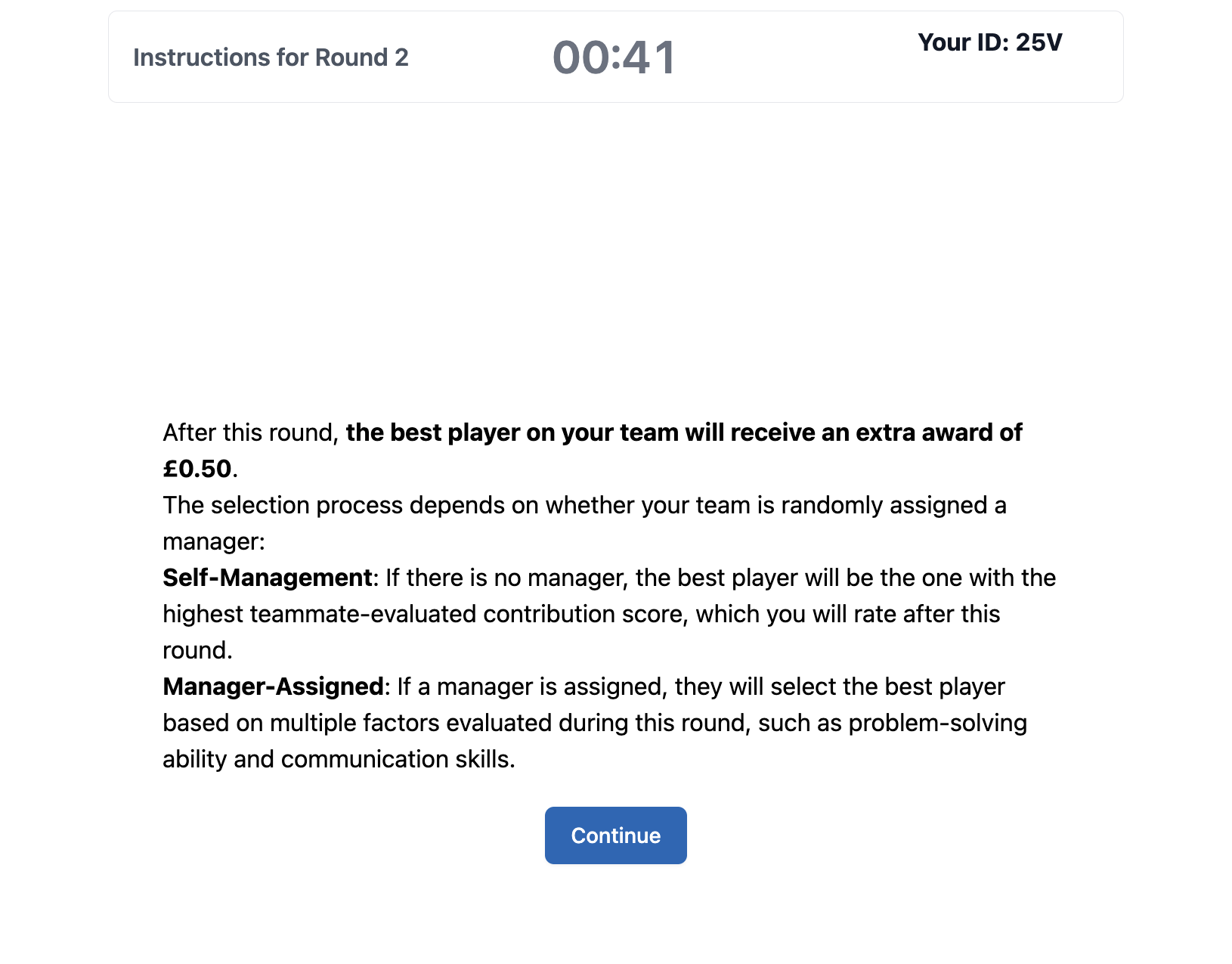} 
		\caption{Instructions for the second round - page 2.} 
		\label{fig:page3} 
	\end{figure} 
	
	\begin{figure}[htbp] 
		\centering
		\includegraphics[width=\textwidth]{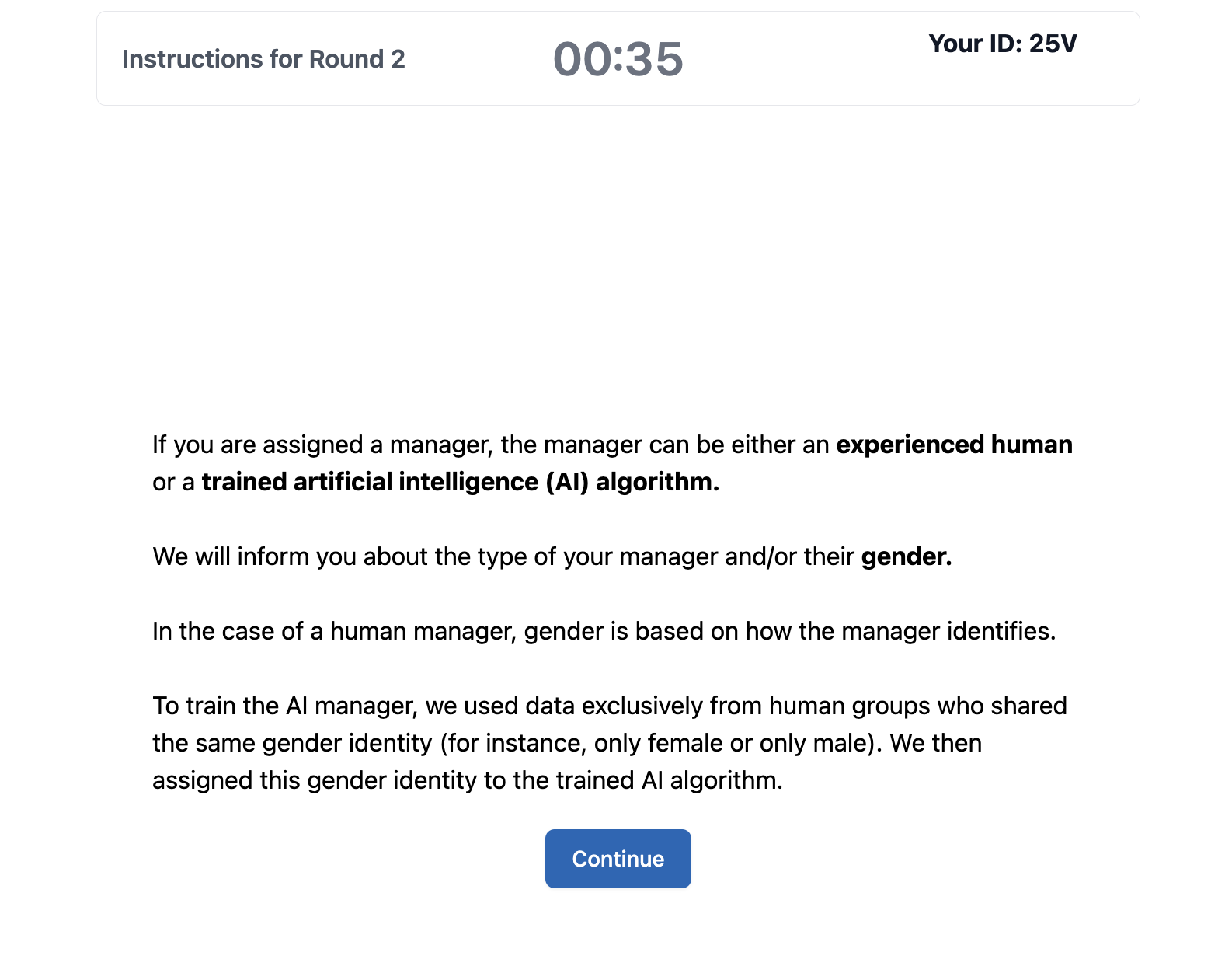} 
		\caption{Instructions for the second round - page 3.} 
		\label{fig:page4} 
	\end{figure} 
	
	\begin{figure}[htbp] 
		\centering
		\includegraphics[width=\textwidth]{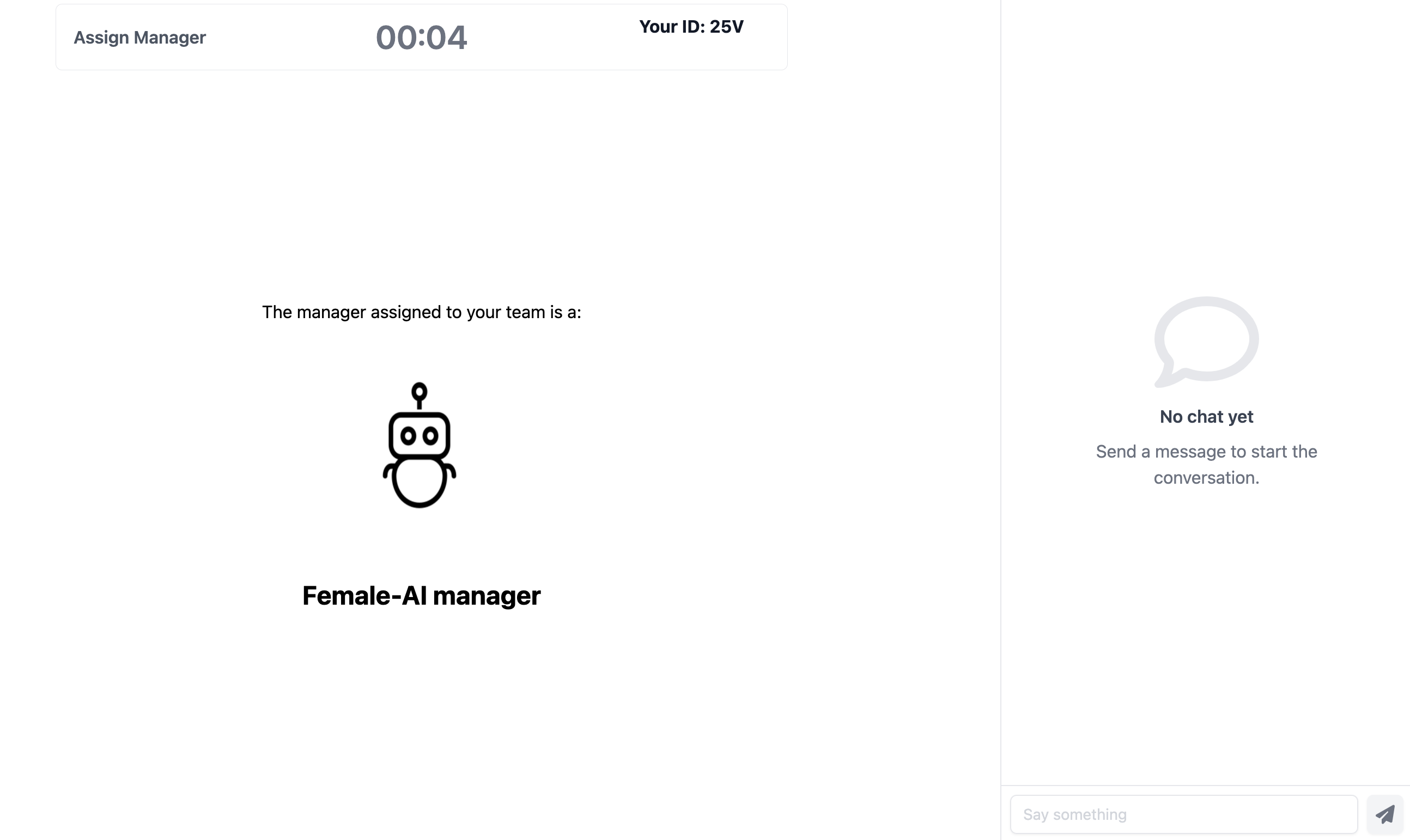} 
		\caption{A manager is randomly assigned to each team. The chat interface allows participants to communicate with their teammates.} 
		\label{fig:page5} 
	\end{figure} 
	
	\begin{figure}[htbp] 
		\centering
		\includegraphics[width=\textwidth]{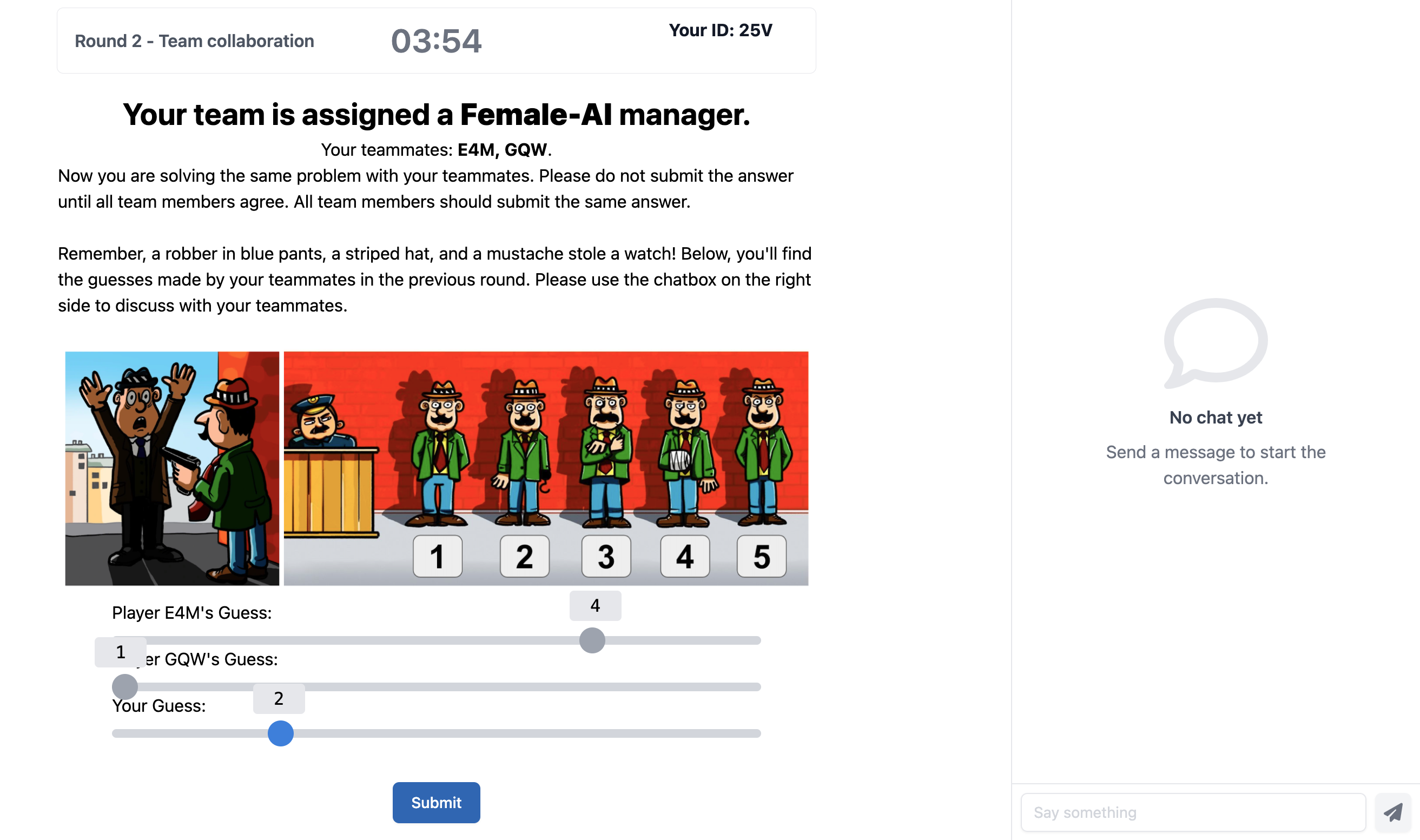} 
		\caption{User interface for the second round. Participants can view their own and their teammates’ responses from the previous round.} 
		\label{fig:page6} 
	\end{figure} 
	
	\begin{figure}[htbp] 
		\centering
		\includegraphics[width=\textwidth]{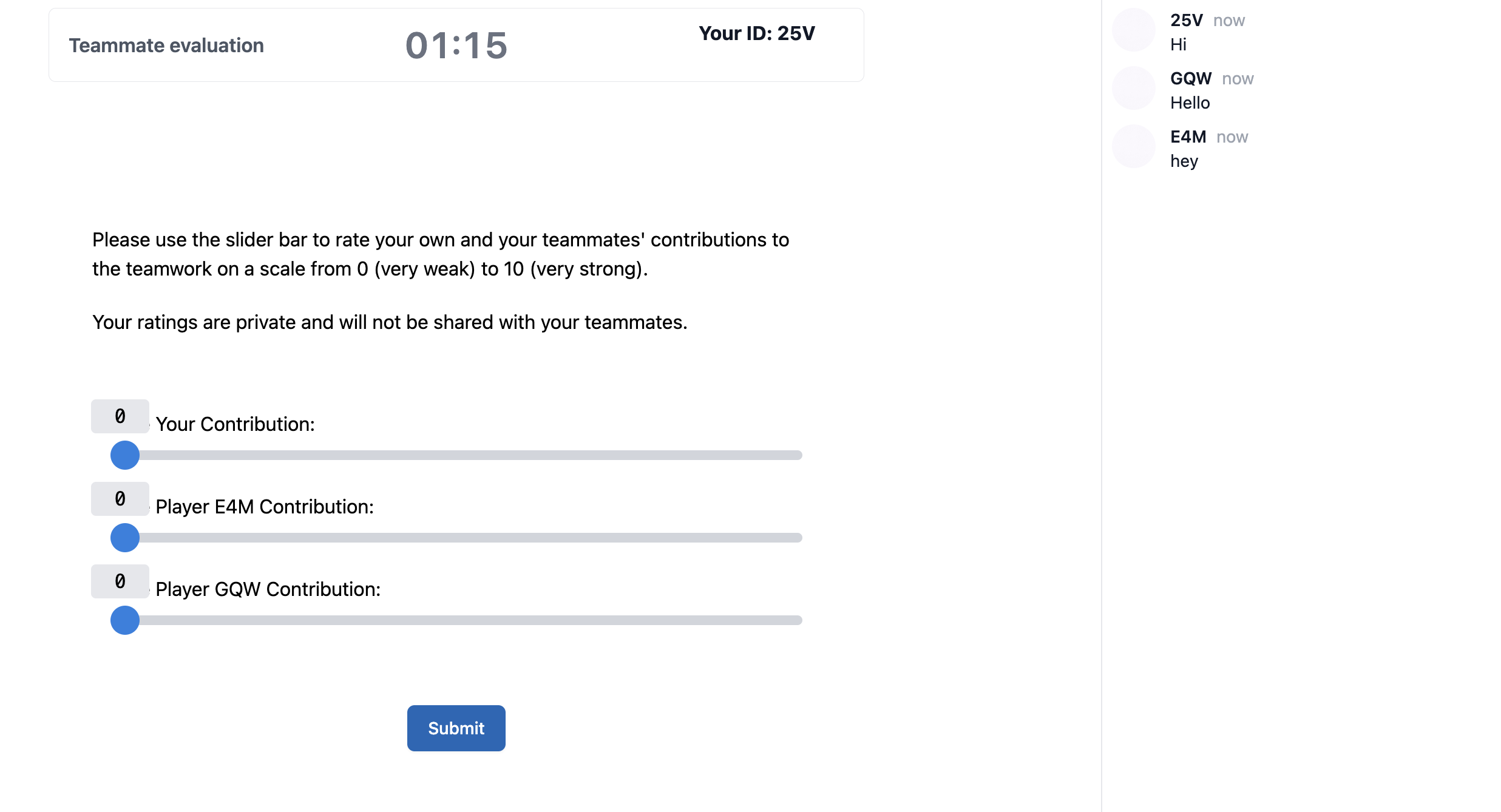} 
		\caption{Participants rate their own and their teammates’ contributions. The chatbox is disabled, allowing them to view past messages but preventing further input.} 
		\label{fig:page7} 
	\end{figure} 
	
	\begin{figure}[htbp] 
		\centering
		\includegraphics[width=\textwidth]{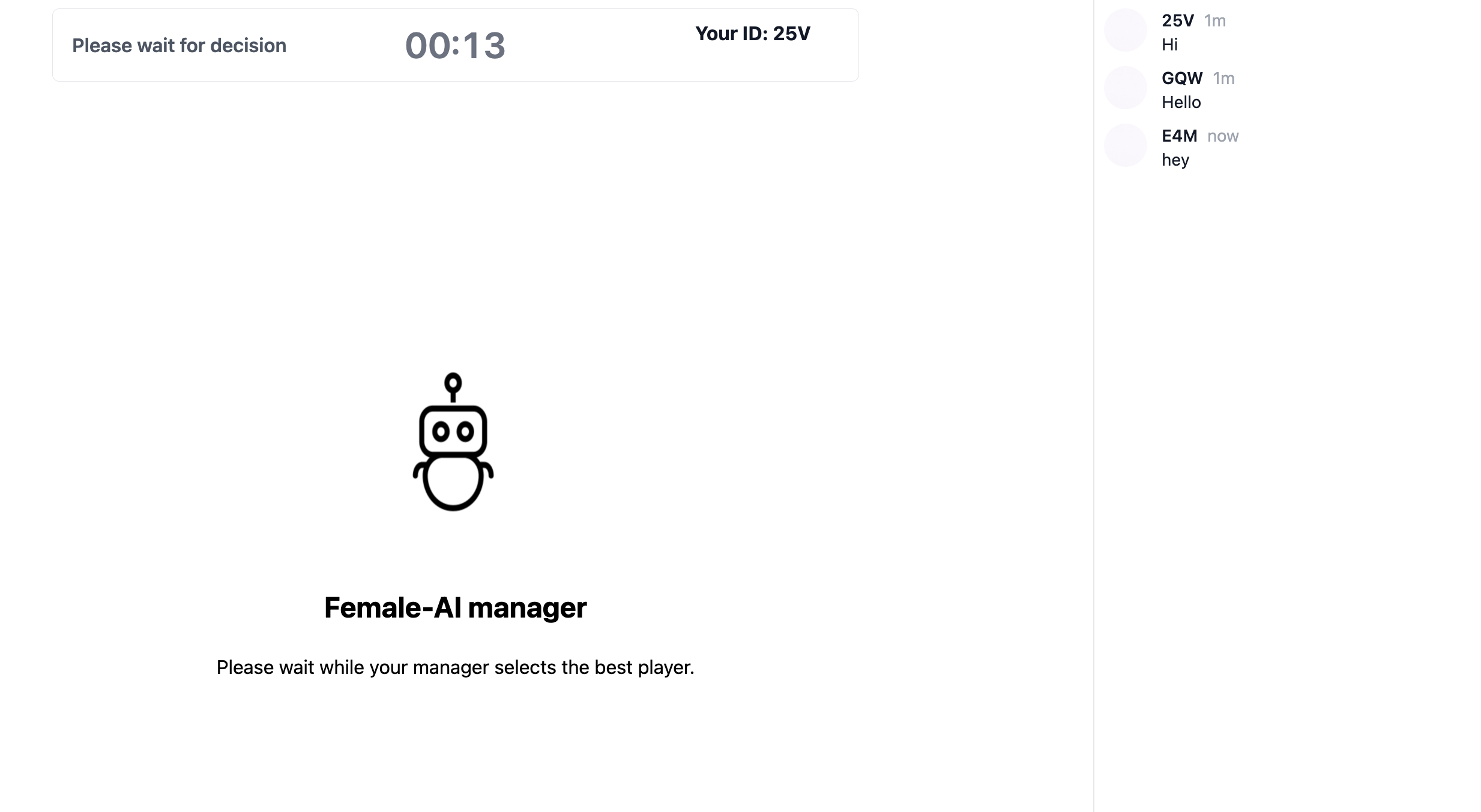} 
		\caption{Participants wait for the manager's decision. The manager's type and gender are displayed again to reinforce memory.} 
		\label{fig:page8} 
	\end{figure} 
	
	\begin{figure}[htbp] 
		\centering
		\includegraphics[width=\textwidth]{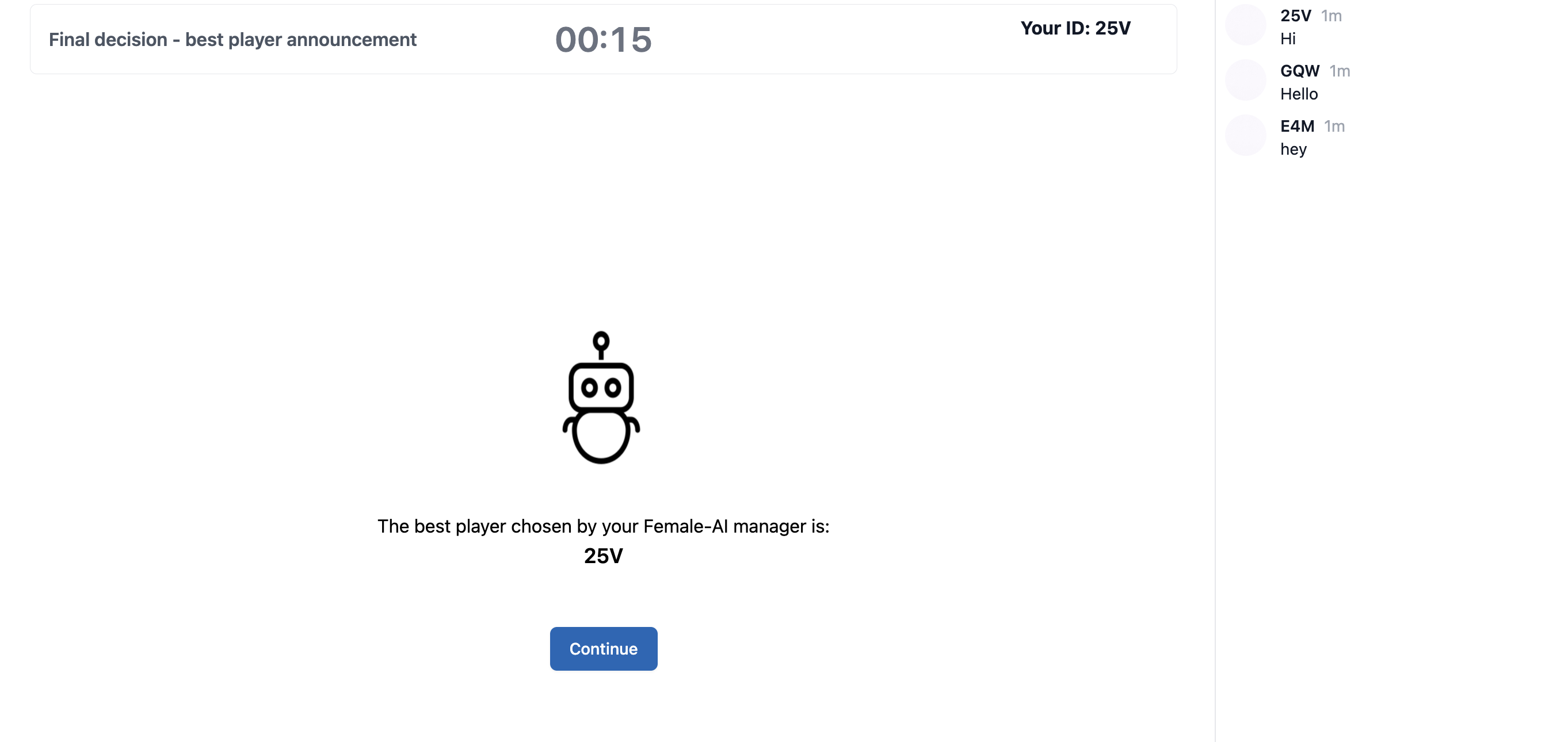} 
		\caption{The manager announces the selected player.} 
		\label{fig:page9} 
	\end{figure} 
	
	\clearpage
	
	\section*{Demographics of participants}
	
	\begin{figure}[htbp] 
		\centering
		\includegraphics[width=\textwidth]{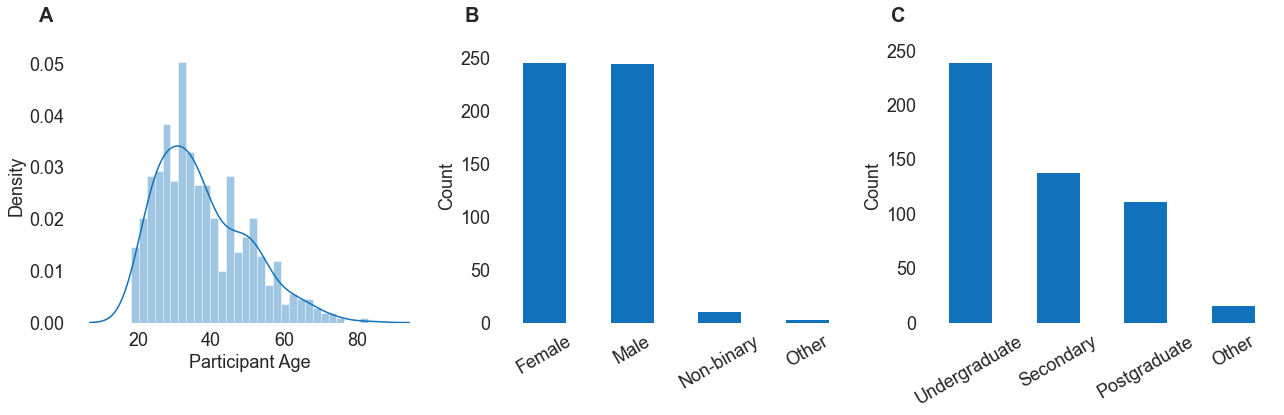} 
		\caption{
			Demographic information of the participants.} 
		\label{fig:demographics} 
	\end{figure} 
	
	\section*{Correlation between variables}
	
	\begin{figure}[htbp] 
		\centering
		\includegraphics[width=\textwidth]{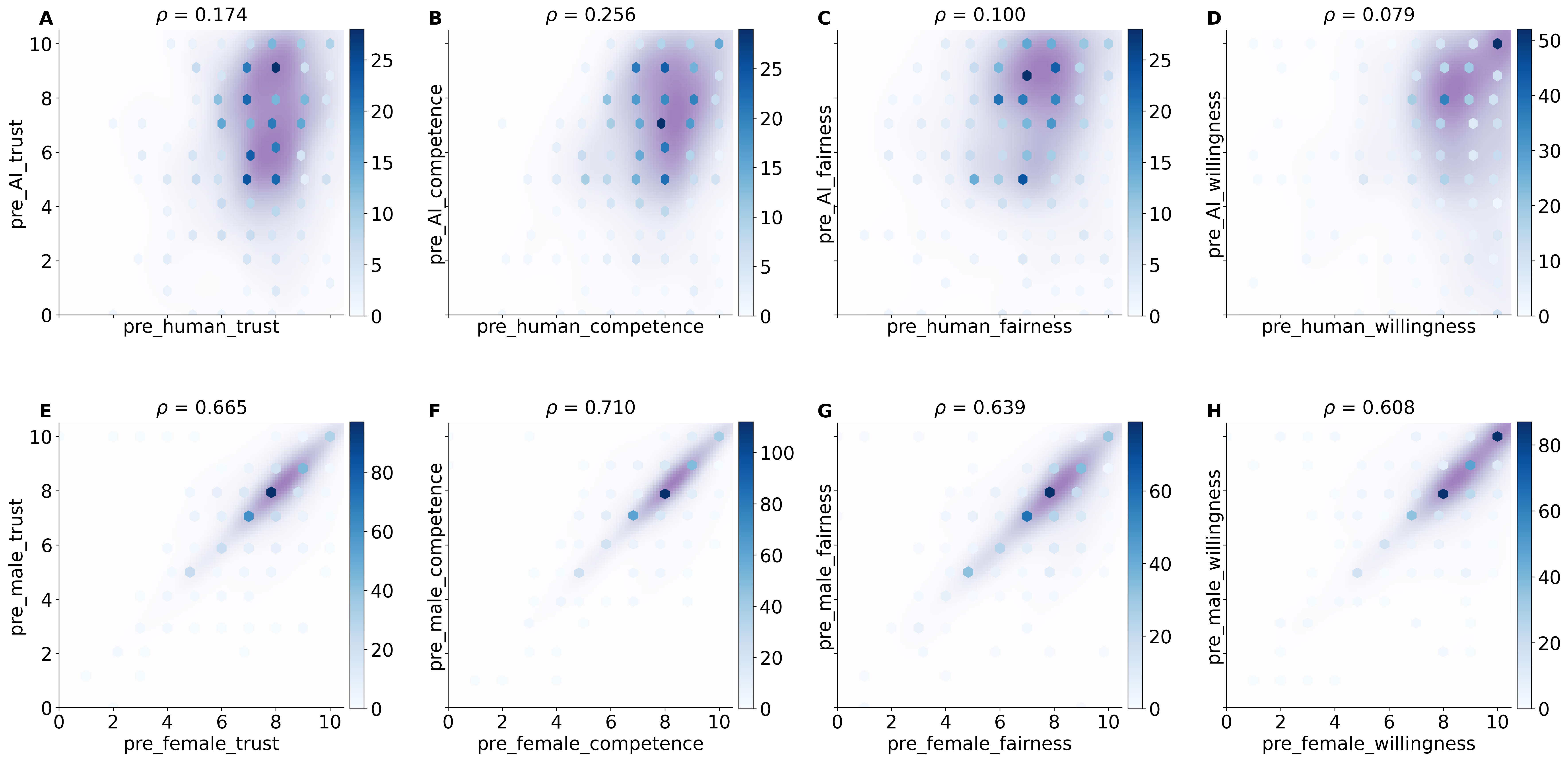} 
		\caption{
			Hexbin plot~\cite{matplotlib_hexbin} of perceived trustworthiness, competence, fairness, and willingness in the pre-treatment survey by manager type (A, B, C, D) and gender (E, F, G, H), with smoother background shading using a 2D Gaussian kernel density estimation (KDE)~\cite{scipy_gaussian_kde} overlay beneath the hexbin plot to enhance the overall visualization.} 
		\label{fig:correlation} 
	\end{figure}

	

	\clearpage
	\section*{Linear regression model results}

	\begin{table}[ht]
		\centering
		\footnotesize	
		\caption{Regression Coefficients Predicting Post-Fairness -- Female Participants}
		\begin{tabular}{lcccc}
			\toprule
			\textbf{Variable} & \textbf{Estimate} & \textbf{Std. Error} & \textbf{t value} & 
			$\mathbf{\text{Pr}(|t|)}$
			\\
			\midrule
			(Intercept) & 4.0395  & 1.4999  & 2.693  & 0.0076 **  \\
			Pre-Type Fairness & 0.1461  & 0.0969  & 1.507  & 0.1331  \\
			Pre-Gender Fairness & 0.3886  & 0.1121  & 3.467  & 0.0006 ***  \\
			Age & -0.0017  & 0.0148  & -0.114  & 0.9092  \\
			Education: Postgraduate & 0.8362  & 1.0070  & 0.830  & 0.4072  \\
			Education: Secondary & 0.6138  & 0.9906  & 0.620  & 0.5361  \\
			Education: Undergraduate & 0.8177  & 0.9672  & 0.845  & 0.3988  \\
			Awarded No: AI Manager: Female & -3.4673  & 0.9257  & -3.745  & 0.0002 ***  \\
			Awarded Yes: AI Manager: Female & -1.1318  & 1.0900  & -1.038  & 0.3002  \\
			Awarded No: Human Manager: Female & -2.2088  & 0.9139  & -2.417  & 0.0164 *  \\
			Awarded Yes: Human Manager: Female & -1.6496  & 1.1124  & -1.483  & 0.1395  \\
			Awarded No: AI Manager: Unspecified & -3.5988  & 0.9294  & -3.872  & 0.0001 ***  \\
			Awarded Yes: AI Manager: Unspecified & -0.4074  & 1.0170  & -0.401  & 0.6891  \\
			Awarded No: Human Manager: Unspecified & -2.3364  & 0.8898  & -2.626  & 0.0092 **  \\
			Awarded Yes: Human Manager: Unspecified & -0.2315  & 1.0433  & -0.222  & 0.8246  \\
			Awarded No: AI Manager: Male & -1.3258  & 0.9172  & -1.445  & 0.1497  \\
			Awarded Yes: AI Manager: Male & 0.6006  & 1.0915  & 0.550  & 0.5827  \\
			Awarded No: Human Manager: Male & -1.6446  & 0.8841  & -1.860  & 0.0642 .  \\
			Awarded Yes: Human Manager: Male & NA  & NA  & NA  & NA  \\
			Awarded No: AI Contribution & -0.4370  & 0.1221  & -3.579  & 0.0004 ***  \\
			Awarded Yes: AI Contribution & 0.0780  & 0.1491  & 0.523  & 0.6014  \\
			Awarded No: Human Contribution & -0.7488  & 0.0960  & -7.799  & 2.34e-13 ***  \\
			Awarded Yes: Human Contribution & 0.0958  & 0.1280  & 0.749  & 0.4549  \\
			\midrule
			\multicolumn{5}{l}{\textbf{\textit{Significance codes:}} 0 ‘***’ 0.001 ‘**’ 0.01 ‘*’ 0.05 ‘.’ 0.1 ‘ ’ 1} \\
			\midrule
			\textbf{\textit{Residuals}} & \multicolumn{4}{c}{} \\
			Min & -7.6668 & 1Q & -1.3399 &  \\
			Median & 0.2766 & 3Q & 1.7927 &  \\
			Max & 7.7341 &  &  &  \\
			\midrule
			\textbf{\textit{Model Summary}}\\
			Residual Standard Error & \multicolumn{4}{c}{2.817 on 224 degrees of freedom} \\
			Multiple R-Squared & \multicolumn{4}{c}{0.4443} \\
			Adjusted R-Squared & \multicolumn{4}{c}{0.3922} \\
			F-statistic & \multicolumn{4}{c}{8.527 on 21 and 224 DF, p $<$ 2.2e-16} \\
			\bottomrule
		\end{tabular}
		\label{tab:regression_fairness_female}
	\end{table}

	\begin{table}[ht]
		\centering
		\footnotesize	
		\caption{Regression Coefficients Predicting Post-Fairness -- Male Participants}
		\begin{tabular}{lcccc}
			\toprule
			\textbf{Variable} & \textbf{Estimate} & \textbf{Std. Error} & \textbf{t value} & 
			$\mathbf{\text{Pr}(|t|)}$
			\\
			\midrule
			(Intercept) & 3.3169  & 1.5544  & 2.134  & 0.0339 *  \\
			Pre-Type Fairness & 0.2945  & 0.0795  & 3.705  & 0.0003 ***  \\
			Pre-Gender Fairness & 0.1235  & 0.0967  & 1.277  & 0.2030  \\
			Age & 0.0074  & 0.0138  & 0.532  & 0.5951  \\
			Education: Postgraduate & 2.3293  & 1.1718  & 1.988  & 0.0481 *  \\
			Education: Secondary & 1.6512  & 1.1618  & 1.421  & 0.1566  \\
			Education: Undergraduate & 1.8422  & 1.1474  & 1.606  & 0.1098  \\
			Awarded No: AI Manager: Female & -2.8713  & 0.8316  & -3.453  & 0.0007 ***  \\
			Awarded Yes: AI Manager: Female & -0.1603  & 0.9657  & -0.166  & 0.8683  \\
			Awarded No: Human Manager: Female & -2.5743  & 0.8496  & -3.030  & 0.0027 **  \\
			Awarded Yes: Human Manager: Female & -0.0991  & 0.9307  & -0.107  & 0.9153  \\
			Awarded No: AI Manager: Unspecified & -2.0047  & 0.8682  & -2.309  & 0.0219 *  \\
			Awarded Yes: AI Manager: Unspecified & -1.0692  & 1.1166  & -0.958  & 0.3393  \\
			Awarded No: Human Manager: Unspecified & -0.7078  & 0.9531  & -0.743  & 0.4585  \\
			Awarded Yes: Human Manager: Unspecified & -0.8012  & 0.9276  & -0.864  & 0.3886  \\
			Awarded No: AI Manager: Male & -1.3576  & 0.8163  & -1.663  & 0.0977 .  \\
			Awarded Yes: AI Manager: Male & -0.6350  & 0.9667  & -0.657  & 0.5120  \\
			Awarded No: Human Manager: Male & -1.4590  & 0.8100  & -1.801  & 0.0730 .  \\
			Awarded Yes: Human Manager: Male & NA  & NA  & NA  & NA  \\
			Awarded No: AI Contribution & -0.4360  & 0.0819  & -5.326  & 2.44e-07 ***  \\
			Awarded Yes: AI Contribution & 0.3264  & 0.1771  & 1.843  & 0.0667 .  \\
			Awarded No: Human Contribution & -0.3763  & 0.1080  & -3.485  & 0.0006 ***  \\
			Awarded Yes: Human Contribution & 0.1369  & 0.1304  & 1.050  & 0.2949  \\
			\midrule
			\multicolumn{5}{l}{\textbf{\textit{Significance codes:}} 0 ‘***’ 0.001 ‘**’ 0.01 ‘*’ 0.05 ‘.’ 0.1 ‘ ’ 1} \\
			\midrule
			\textbf{\textit{Residuals}} & \multicolumn{4}{c}{} \\
			Min & -6.4840 & 1Q & -1.2001 &  \\
			Median & 0.4108 & 3Q & 1.6587 &  \\
			Max & 5.3389 &  &  &  \\
			\midrule
			\textbf{\textit{Model Summary}}\\
			Residual Standard Error & \multicolumn{4}{c}{2.428 on 223 degrees of freedom} \\
			Multiple R-Squared & \multicolumn{4}{c}{0.4087} \\
			Adjusted R-Squared & \multicolumn{4}{c}{0.3530} \\
			F-statistic & \multicolumn{4}{c}{7.339 on 21 and 223 DF, p = 3.203e-16} \\
			\bottomrule
		\end{tabular}
		\label{tab:regression_fairness_male}
	\end{table}

	\begin{table}[ht]
		\centering
		\footnotesize	
		\caption{Regression Coefficients Predicting Post-Trust -- Female Participants}
		\begin{tabular}{lcccc}
			\toprule
			\textbf{Variable} & \textbf{Estimate} & \textbf{Std. Error} & \textbf{t value} & 
			$\mathbf{\text{Pr}(|t|)}$
			\\
			\midrule
			(Intercept) & 3.4743  & 1.4814  & 2.345  & 0.0199 *  \\
			Pre-Type Trust & 0.2121  & 0.1106  & 1.918  & 0.0564 .  \\
			Pre-Gender Trust & 0.3450  & 0.1177  & 2.930  & 0.0037 **  \\
			Age & 0.0012  & 0.0140  & 0.085  & 0.9323  \\
			Education: Postgraduate & 1.0122  & 0.9533  & 1.062  & 0.2894  \\
			Education: Secondary & 0.5899  & 0.9361  & 0.630  & 0.5292  \\
			Education: Undergraduate & 1.0808  & 0.9121  & 1.185  & 0.2373  \\
			Awarded No: AI Manager: Female & -3.1587  & 0.8771  & -3.601  & 0.0004 ***  \\
			Awarded Yes: AI Manager: Female & -0.2714  & 1.0573  & -0.257  & 0.7977  \\
			Awarded No: Human Manager: Female & -1.9109  & 0.8643  & -2.211  & 0.0280 *  \\
			Awarded Yes: Human Manager: Female & -1.4303  & 1.0422  & -1.372  & 0.1713  \\
			Awarded No: AI Manager: Unspecified & -3.0220  & 0.8974  & -3.367  & 0.0009 ***  \\
			Awarded Yes: AI Manager: Unspecified & -0.0905  & 0.9708  & -0.093  & 0.9258  \\
			Awarded No: Human Manager: Unspecified & -2.0843  & 0.8383  & -2.486  & 0.0136 *  \\
			Awarded Yes: Human Manager: Unspecified & -0.0701  & 0.9796  & -0.072  & 0.9430  \\
			Awarded No: AI Manager: Male & -1.3278  & 0.8746  & -1.518  & 0.1304  \\
			Awarded Yes: AI Manager: Male & 0.6407  & 1.0306  & 0.622  & 0.5348  \\
			Awarded No: Human Manager: Male & -1.7224  & 0.8383  & -2.055  & 0.0411 *  \\
			Awarded Yes: Human Manager: Male & NA  & NA  & NA  & NA  \\
			Awarded No: AI Contribution & -0.4236  & 0.1154  & -3.672  & 0.0003 ***  \\
			Awarded Yes: AI Contribution & 0.0482  & 0.1402  & 0.344  & 0.7311  \\
			Awarded No: Human Contribution & -0.7140  & 0.0898  & -7.955  & 8.85e-14 ***  \\
			Awarded Yes: Human Contribution & 0.0840  & 0.1209  & 0.695  & 0.4880  \\
			\midrule
			\multicolumn{5}{l}{\textbf{\textit{Significance codes:}} 0 ‘***’ 0.001 ‘**’ 0.01 ‘*’ 0.05 ‘.’ 0.1 ‘ ’ 1} \\
			\midrule
			\textbf{\textit{Residuals}} & \multicolumn{4}{c}{} \\
			Min & -7.5758 & 1Q & -1.3805 &  \\
			Median & 0.3509 & 3Q & 1.6970 &  \\
			Max & 7.2637 &  &  &  \\
			\midrule
			\textbf{\textit{Model Summary}}\\
			Residual Standard Error & \multicolumn{4}{c}{2.655 on 224 degrees of freedom} \\
			Multiple R-Squared & \multicolumn{4}{c}{0.4464} \\
			Adjusted R-Squared & \multicolumn{4}{c}{0.3945} \\
			F-statistic & \multicolumn{4}{c}{8.601 on 21 and 224 DF, p $<$ 2.2e-16} \\
			\bottomrule
		\end{tabular}
		\label{tab:regression_trust_female}
	\end{table}

	\begin{table}[ht]
		\centering
		\footnotesize	
		\caption{Regression Coefficients Predicting Post-Trust -- Male Participants}
		\begin{tabular}{lcccc}
			\toprule
			\textbf{Variable} & \textbf{Estimate} & \textbf{Std. Error} & \textbf{t value} & 
			$\mathbf{\text{Pr}(|t|)}$
			\\
			\midrule
			(Intercept) & 3.6564  & 1.5516  & 2.357  & 0.0193 *  \\
			Pre-Type Trust & 0.2924  & 0.0743  & 3.936  & 0.0001 ***  \\
			Pre-Gender Trust & 0.1887  & 0.1028  & 1.836  & 0.0677 .  \\
			Age & -0.0013  & 0.0132  & -0.097  & 0.9230  \\
			Education: Postgraduate & 1.3421  & 1.1166  & 1.202  & 0.2307  \\
			Education: Secondary & 1.0498  & 1.1080  & 0.947  & 0.3445  \\
			Education: Undergraduate & 1.0350  & 1.0951  & 0.945  & 0.3456  \\
			Awarded No: AI Manager: Female & -2.3691  & 0.8005  & -2.960  & 0.0034 **  \\
			Awarded Yes: AI Manager: Female & 0.2696  & 0.9103  & 0.296  & 0.7674  \\
			Awarded No: Human Manager: Female & -2.1784  & 0.8164  & -2.668  & 0.0082 **  \\
			Awarded Yes: Human Manager: Female & 0.0699  & 0.8923  & 0.078  & 0.9376  \\
			Awarded No: AI Manager: Unspecified & -1.7127  & 0.8296  & -2.064  & 0.0401 *  \\
			Awarded Yes: AI Manager: Unspecified & -0.8845  & 1.0667  & -0.829  & 0.4079  \\
			Awarded No: Human Manager: Unspecified & -0.2661  & 0.9196  & -0.289  & 0.7725  \\
			Awarded Yes: Human Manager: Unspecified & -0.4648  & 0.8914  & -0.521  & 0.6026  \\
			Awarded No: AI Manager: Male & -1.0990  & 0.7809  & -1.407  & 0.1607  \\
			Awarded Yes: AI Manager: Male & -0.0543  & 0.9260  & -0.059  & 0.9533  \\
			Awarded No: Human Manager: Male & -1.3263  & 0.7721  & -1.718  & 0.0872 .  \\
			Awarded Yes: Human Manager: Male & NA  & NA  & NA  & NA  \\
			Awarded No: AI Contribution & -0.4088  & 0.0782  & -5.231  & 3.88e-07 ***  \\
			Awarded Yes: AI Contribution & 0.2895  & 0.1698  & 1.705  & 0.0896 .  \\
			Awarded No: Human Contribution & -0.3741  & 0.1030  & -3.632  & 0.0003 ***  \\
			Awarded Yes: Human Contribution & 0.1524  & 0.1240  & 1.229  & 0.2203  \\
			\midrule
			\multicolumn{5}{l}{\textbf{\textit{Significance codes:}} 0 ‘***’ 0.001 ‘**’ 0.01 ‘*’ 0.05 ‘.’ 0.1 ‘ ’ 1} \\
			\midrule
			\textbf{\textit{Residuals}} & \multicolumn{4}{c}{} \\
			Min & -7.5853 & 1Q & -1.2865 &  \\
			Median & 0.4028 & 3Q & 1.5470 &  \\
			Max & 4.2839 &  &  &  \\
			\midrule
			\textbf{\textit{Model Summary}}\\
			Residual Standard Error & \multicolumn{4}{c}{2.316 on 223 degrees of freedom} \\
			Multiple R-Squared & \multicolumn{4}{c}{0.4074} \\
			Adjusted R-Squared & \multicolumn{4}{c}{0.3516} \\
			F-statistic & \multicolumn{4}{c}{7.302 on 21 and 223 DF, p = 3.934e-16} \\
			\bottomrule
		\end{tabular}
		\label{tab:regression_trust_male}
	\end{table}

	\begin{table}[ht]
		\centering
		\footnotesize	
		\caption{Regression Coefficients Predicting Post-Competence -- Female Participants}
		\begin{tabular}{lcccc}
			\toprule
			\textbf{Variable} & \textbf{Estimate} & \textbf{Std. Error} & \textbf{t value} & 
			$\mathbf{\text{Pr}(|t|)}$
			\\
			\midrule
			(Intercept) & 3.1400  & 1.4832  & 2.117  & 0.0354 *  \\
			Pre-Type Competence & 0.2240  & 0.1040  & 2.154  & 0.0323 *  \\
			Pre-Gender Competence & 0.3483  & 0.1221  & 2.852  & 0.0047 **  \\
			Age & 0.0024  & 0.0140  & 0.170  & 0.8655  \\
			Education: Postgraduate & 0.9337  & 0.9499  & 0.983  & 0.3267  \\
			Education: Secondary & 0.7219  & 0.9373  & 0.770  & 0.4420  \\
			Education: Undergraduate & 1.1632  & 0.9137  & 1.273  & 0.2043  \\
			Awarded No: AI Manager: Female & -2.8639  & 0.8859  & -3.233  & 0.0014 **  \\
			Awarded Yes: AI Manager: Female & -0.1670  & 1.0564  & -0.158  & 0.8746  \\
			Awarded No: Human Manager: Female & -1.9253  & 0.8684  & -2.217  & 0.0276 *  \\
			Awarded Yes: Human Manager: Female & -1.3160  & 1.0512  & -1.252  & 0.2119  \\
			Awarded No: AI Manager: Unspecified & -2.9419  & 0.9102  & -3.232  & 0.0014 **  \\
			Awarded Yes: AI Manager: Unspecified & -0.1315  & 0.9739  & -0.135  & 0.8927  \\
			Awarded No: Human Manager: Unspecified & -2.1825  & 0.8403  & -2.597  & 0.0100 *  \\
			Awarded Yes: Human Manager: Unspecified & -0.0143  & 0.9824  & -0.015  & 0.9884  \\
			Awarded No: AI Manager: Male & -1.2305  & 0.8761  & -1.405  & 0.1615  \\
			Awarded Yes: AI Manager: Male & 0.3554  & 1.0363  & 0.343  & 0.7319  \\
			Awarded No: Human Manager: Male & -1.6042  & 0.8384  & -1.913  & 0.0570 .  \\
			Awarded Yes: Human Manager: Male & NA  & NA  & NA  & NA  \\
			Awarded No: AI Contribution & -0.3881  & 0.1131  & -3.430  & 0.0007 ***  \\
			Awarded Yes: AI Contribution & 0.0400  & 0.1406  & 0.284  & 0.7764  \\
			Awarded No: Human Contribution & -0.7550  & 0.0902  & -8.374  & 6.1e-15 ***  \\
			Awarded Yes: Human Contribution & 0.0977  & 0.1211  & 0.807  & 0.4208  \\
			\midrule
			\multicolumn{5}{l}{\textbf{\textit{Significance codes:}} 0 ‘***’ 0.001 ‘**’ 0.01 ‘*’ 0.05 ‘.’ 0.1 ‘ ’ 1} \\
			\midrule
			\textbf{\textit{Residuals}} & \multicolumn{4}{c}{} \\
			Min & -7.7431 & 1Q & -1.5865 &  \\
			Median & 0.3368 & 3Q & 1.7457 &  \\
			Max & 7.1707 &  &  &  \\
			\midrule
			\textbf{\textit{Model Summary}}\\
			Residual Standard Error & \multicolumn{4}{c}{2.662 on 224 degrees of freedom} \\
			Multiple R-Squared & \multicolumn{4}{c}{0.4525} \\
			Adjusted R-Squared & \multicolumn{4}{c}{0.4012} \\
			F-statistic & \multicolumn{4}{c}{8.817 on 21 and 224 DF, p $<$ 2.2e-16} \\
			\bottomrule
		\end{tabular}
		\label{tab:regression_competence_female}
	\end{table}

	\begin{table}[ht]
		\centering
		\footnotesize	
		\caption{Regression Coefficients Predicting Post-Competence -- Male Participants}
		\begin{tabular}{lcccc}
			\toprule
			\textbf{Variable} & \textbf{Estimate} & \textbf{Std. Error} & \textbf{t value} & 
			$\mathbf{\text{Pr}(|t|)}$
			\\
			\midrule
			(Intercept) & 3.2565  & 1.5645  & 2.082  & 0.0385 *  \\
			Pre-Type Competence & 0.3775  & 0.0777  & 4.858  & 2.23e-06 ***  \\
			Pre-Gender Competence & 0.0748  & 0.1047  & 0.715  & 0.4753  \\
			Age & 0.0022  & 0.0135  & 0.164  & 0.8697  \\
			Education: Postgraduate & 2.3987  & 1.1421  & 2.100  & 0.0368 *  \\
			Education: Secondary & 1.6722  & 1.1328  & 1.476  & 0.1413  \\
			Education: Undergraduate & 1.9537  & 1.1191  & 1.746  & 0.0822 .  \\
			Awarded No: AI Manager: Female & -2.7103  & 0.8198  & -3.306  & 0.0011 **  \\
			Awarded Yes: AI Manager: Female & 0.0795  & 0.9345  & 0.085  & 0.9323  \\
			Awarded No: Human Manager: Female & -2.8066  & 0.8280  & -3.390  & 0.0008 ***  \\
			Awarded Yes: Human Manager: Female & -0.2948  & 0.9067  & -0.325  & 0.7454  \\
			Awarded No: AI Manager: Unspecified & -2.3198  & 0.8494  & -2.731  & 0.0068 **  \\
			Awarded Yes: AI Manager: Unspecified & -0.5771  & 1.0888  & -0.530  & 0.5966  \\
			Awarded No: Human Manager: Unspecified & -1.0009  & 0.9321  & -1.074  & 0.2841  \\
			Awarded Yes: Human Manager: Unspecified & -1.1244  & 0.9066  & -1.240  & 0.2162  \\
			Awarded No: AI Manager: Male & -1.4461  & 0.8038  & -1.799  & 0.0734 .  \\
			Awarded Yes: AI Manager: Male & -0.2052  & 0.9516  & -0.216  & 0.8295  \\
			Awarded No: Human Manager: Male & -1.6669  & 0.7888  & -2.113  & 0.0357 *  \\
			Awarded Yes: Human Manager: Male & NA  & NA  & NA  & NA  \\
			Awarded No: AI Contribution & -0.4253  & 0.0801  & -5.311  & 2.63e-07 ***  \\
			Awarded Yes: AI Contribution & 0.2262  & 0.1732  & 1.306  & 0.1929  \\
			Awarded No: Human Contribution & -0.3799  & 0.1055  & -3.602  & 0.0004 ***  \\
			Awarded Yes: Human Contribution & 0.1181  & 0.1273  & 0.928  & 0.3545  \\
			\midrule
			\multicolumn{5}{l}{\textbf{\textit{Significance codes:}} 0 ‘***’ 0.001 ‘**’ 0.01 ‘*’ 0.05 ‘.’ 0.1 ‘ ’ 1} \\
			\midrule
			\textbf{\textit{Residuals}} & \multicolumn{4}{c}{} \\
			Min & -6.2809 & 1Q & -1.3187 &  \\
			Median & 0.3821 & 3Q & 1.7312 &  \\
			Max & 4.5079 &  &  &  \\
			\midrule
			\textbf{\textit{Model Summary}}\\
			Residual Standard Error & \multicolumn{4}{c}{2.368 on 223 degrees of freedom} \\
			Multiple R-Squared & \multicolumn{4}{c}{0.437} \\
			Adjusted R-Squared & \multicolumn{4}{c}{0.3839} \\
			F-statistic & \multicolumn{4}{c}{8.241 on 21 and 223 DF, p $<$ 2.2e-16} \\
			\bottomrule
		\end{tabular}
		\label{tab:regression_competence_male}
	\end{table}

	\begin{table}[ht]
		\centering
		\footnotesize	
		\caption{Regression Coefficients Predicting Post-Willingness -- Female Participants}
		\begin{tabular}{lcccc}
			\toprule
			\textbf{Variable} & \textbf{Estimate} & \textbf{Std. Error} & \textbf{t value} & 
			$\mathbf{\text{Pr}(|t|)}$
			\\
			\midrule
			(Intercept) & 4.4705  & 1.7351  & 2.577  & 0.0106 *  \\
			Pre-Type Willingness & 0.1704  & 0.1098  & 1.552  & 0.1221  \\
			Pre-Gender Willingness & 0.2132  & 0.1319  & 1.616  & 0.1074  \\
			Age & 0.0051  & 0.0158  & 0.321  & 0.7486  \\
			Education: Postgraduate & 0.7707  & 1.0733  & 0.718  & 0.4735  \\
			Education: Secondary & 0.4296  & 1.0598  & 0.405  & 0.6856  \\
			Education: Undergraduate & 0.7251  & 1.0322  & 0.702  & 0.4831  \\
			Awarded No: AI Manager: Female & -2.9796  & 1.0028  & -2.971  & 0.0033 **  \\
			Awarded Yes: AI Manager: Female & -0.1411  & 1.1818  & -0.119  & 0.9051  \\
			Awarded No: Human Manager: Female & -2.0622  & 0.9861  & -2.091  & 0.0376 *  \\
			Awarded Yes: Human Manager: Female & -0.8737  & 1.1850  & -0.737  & 0.4617  \\
			Awarded No: AI Manager: Unspecified & -2.6885  & 1.0642  & -2.526  & 0.0122 *  \\
			Awarded Yes: AI Manager: Unspecified & -0.4021  & 1.1027  & -0.365  & 0.7157  \\
			Awarded No: Human Manager: Unspecified & -2.0250  & 0.9528  & -2.125  & 0.0347 *  \\
			Awarded Yes: Human Manager: Unspecified & 0.1832  & 1.1124  & 0.165  & 0.8694  \\
			Awarded No: AI Manager: Male & -1.5046  & 0.9917  & -1.517  & 0.1306  \\
			Awarded Yes: AI Manager: Male & -0.0472  & 1.1821  & -0.040  & 0.9682  \\
			Awarded No: Human Manager: Male & -1.3881  & 0.9608  & -1.445  & 0.1499  \\
			Awarded Yes: Human Manager: Male & NA  & NA  & NA  & NA  \\
			Awarded No: AI Contribution & -0.4318  & 0.1300  & -3.322  & 0.0010 **  \\
			Awarded Yes: AI Contribution & 0.0730  & 0.1600  & 0.456  & 0.6487  \\
			Awarded No: Human Contribution & -0.6829  & 0.1020  & -6.693  & 1.74e-10 ***  \\
			Awarded Yes: Human Contribution & 0.0502  & 0.1373  & 0.366  & 0.7147  \\
			\midrule
			\multicolumn{5}{l}{\textbf{\textit{Significance codes:}} 0 ‘***’ 0.001 ‘**’ 0.01 ‘*’ 0.05 ‘.’ 0.1 ‘ ’ 1} \\
			\midrule
			\textbf{\textit{Residuals}} & \multicolumn{4}{c}{} \\
			Min & -9.0410 & 1Q & -1.6230 &  \\
			Median & 0.5900 & 3Q & 1.8280 &  \\
			Max & 8.1370 &  &  &  \\
			\midrule
			\textbf{\textit{Model Summary}}\\
			Residual Standard Error & \multicolumn{4}{c}{3.015 on 224 degrees of freedom} \\
			Multiple R-Squared & \multicolumn{4}{c}{0.3624} \\
			Adjusted R-Squared & \multicolumn{4}{c}{0.3026} \\
			F-statistic & \multicolumn{4}{c}{6.062 on 21 and 224 DF, p = 3.919e-13} \\
			\bottomrule
		\end{tabular}
		\label{tab:regression_willingness_female}
	\end{table}

	\begin{table}[ht]
		\centering
		\footnotesize	
		\caption{Regression Coefficients Predicting Post-Willingness -- Male Participants}
		\begin{tabular}{lcccc}
			\toprule
			\textbf{Variable} & \textbf{Estimate} & \textbf{Std. Error} & \textbf{t value} & 
			$\mathbf{\text{Pr}(|t|)}$
			\\
			\midrule
			(Intercept) & 5.3938  & 1.6947  & 3.183  & 0.0017 **  \\
			Pre-Type Willingness & 0.3687  & 0.0688  & 5.361  & 2.06e-07 ***  \\
			Pre-Gender Willingness & -0.0986  & 0.0951  & -1.037  & 0.3006  \\
			Age & -0.0038  & 0.0150  & -0.251  & 0.8020  \\
			Education: Postgraduate & 2.3230  & 1.2620  & 1.841  & 0.0670 .  \\
			Education: Secondary & 1.4247  & 1.2537  & 1.136  & 0.2570  \\
			Education: Undergraduate & 1.5182  & 1.2380  & 1.226  & 0.2214  \\
			Awarded No: AI Manager: Female & -2.8736  & 0.8951  & -3.210  & 0.0015 **  \\
			Awarded Yes: AI Manager: Female & 0.4109  & 1.0362  & 0.397  & 0.6921  \\
			Awarded No: Human Manager: Female & -3.0508  & 0.9151  & -3.334  & 0.0010 **  \\
			Awarded Yes: Human Manager: Female & -0.4271  & 1.0020  & -0.426  & 0.6703  \\
			Awarded No: AI Manager: Unspecified & -1.9181  & 0.9430  & -2.034  & 0.0431 *  \\
			Awarded Yes: AI Manager: Unspecified & -0.5242  & 1.2046  & -0.435  & 0.6639  \\
			Awarded No: Human Manager: Unspecified & -1.7413  & 1.0307  & -1.689  & 0.0925 .  \\
			Awarded Yes: Human Manager: Unspecified & -1.1603  & 1.0054  & -1.154  & 0.2497  \\
			Awarded No: AI Manager: Male & -1.6411  & 0.8810  & -1.863  & 0.0638 .  \\
			Awarded Yes: AI Manager: Male & -0.2509  & 1.0462  & -0.240  & 0.8107  \\
			Awarded No: Human Manager: Male & -1.8619  & 0.8720  & -2.135  & 0.0338 *  \\
			Awarded Yes: Human Manager: Male & NA  & NA  & NA  & NA  \\
			Awarded No: AI Contribution & -0.4318  & 0.0880  & -4.906  & 1.79e-06 ***  \\
			Awarded Yes: AI Contribution & -0.0776  & 0.1881  & -0.412  & 0.6805  \\
			Awarded No: Human Contribution & -0.4337  & 0.1163  & -3.729  & 0.0002 ***  \\
			Awarded Yes: Human Contribution & 0.0531  & 0.1402  & 0.379  & 0.7051  \\
			\midrule
			\multicolumn{5}{l}{\textbf{\textit{Significance codes:}} 0 ‘***’ 0.001 ‘**’ 0.01 ‘*’ 0.05 ‘.’ 0.1 ‘ ’ 1} \\
			\midrule
			\textbf{\textit{Residuals}} & \multicolumn{4}{c}{} \\
			Min & -7.4771 & 1Q & -1.0980 &  \\
			Median & 0.4345 & 3Q & 1.7196 &  \\
			Max & 4.9272 &  &  &  \\
			\midrule
			\textbf{\textit{Model Summary}}\\
			Residual Standard Error & \multicolumn{4}{c}{2.617 on 223 degrees of freedom} \\
			Multiple R-Squared & \multicolumn{4}{c}{0.3807} \\
			Adjusted R-Squared & \multicolumn{4}{c}{0.3224} \\
			F-statistic & \multicolumn{4}{c}{6.527 on 21 and 223 DF, p = 2.883e-14} \\
			\bottomrule
		\end{tabular}
		\label{tab:regression_willingness_male}
	\end{table}

\end{document}